\documentclass[reprint,
superscriptaddress,showpacs,preprintnumbers,
nofootinbib,amsmath,amssymb, aps, prl,longbibliography, twocolumn]{revtex4-2}

\usepackage[utf8]{inputenc}
\usepackage{mathrsfs}
\usepackage{bm}
\usepackage{url}
\usepackage[normalem]{ulem}
\usepackage{mathtools}
\usepackage{array}
\newcolumntype{P}[1]{>{\centering\arraybackslash}p{#1}}
\newcolumntype{M}[1]{>{\centering\arraybackslash}m{#1}}
\usepackage[caption=false]{subfig}
\usepackage{dcolumn}
\usepackage{graphicx, epsfig}
\usepackage[dvipsnames]{xcolor}
\usepackage{mathrsfs}
\usepackage{bm}
\usepackage{yhmath}
\usepackage[caption=false]{subfig}
\usepackage[normalem]{ulem}
\usepackage{mathtools}
\usepackage{bigints}
\usepackage{float}
\usepackage[colorlinks = true, linkcolor = purple, urlcolor  = blue, citecolor = blue, anchorcolor = blue]{hyperref}
\usepackage{float}
\usepackage{multirow}
\usepackage{tikz,xcolor,hyperref}
\definecolor{darkgreen}{rgb}{0.0, 0.2, 0.13}
\definecolor{bostonuniversityred}{rgb}{0.8, 0.0, 0.0}
\definecolor{lime}{HTML}{A6CE39}
\DeclareRobustCommand{\orcidicon}{
	\begin{tikzpicture}
	\draw[lime, fill=lime] (0,0) 
	circle [radius=0.16] 
	node[white] {{\fontfamily{qag}\selectfont \tiny ID}};
	\draw[white, fill=white] (-0.0625,0.095) 
	circle [radius=0.007];
	\end{tikzpicture}
	\hspace{-2mm}
}
\foreach \x in {A, ..., Z}{\expandafter\xdef\csname orcid\x\endcsname{\noexpand\href{https://orcid.org/\csname orcidauthor\x\endcsname}
			{\noexpand\orcidicon}}
}
\usepackage{tikz,xcolor,hyperref}
\definecolor{darkgreen}{rgb}{0.0, 0.2, 0.13}
\definecolor{bostonuniversityred}{rgb}{0.8, 0.0, 0.0}
\definecolor{lime}{HTML}{A6CE39}
\DeclareRobustCommand{\orcidicon}{
	\begin{tikzpicture}
	\draw[lime, fill=lime] (0,0) 
	circle [radius=0.16] 
	node[white] {{\fontfamily{qag}\selectfont \tiny ID}};
	\draw[white, fill=white] (-0.0625,0.095) 
	circle [radius=0.007];
	\end{tikzpicture}
	\hspace{-2mm}
}
\foreach \x in {A, ..., Z}{\expandafter\xdef\csname orcid\x\endcsname{\noexpand\href{https://orcid.org/\csname orcidauthor\x\endcsname}
			{\noexpand\orcidicon}}
}

\def\t13{\mathrel{{\theta_{13}}}}
\def\y12{\mathrel{{\tan^2 \theta_{12}}}}
\def\c2{\mathrel{{\chi^2 }}}

\newcommand{\be}{\begin{equation}}
\newcommand{\ee}{\end{equation}}
\newcommand{\ba}{\begin{eqnarray}}
\newcommand{\ea}{\end{eqnarray}}

\begin{document}
\preprint{\texttt{FERMILAB-PUB-26-0305-T}}
\title{Ultra high-energy cosmic rays from relativistic outflows in accretion induced collapse of white dwarfs}
\author{Mainak Mukhopadhyay\hspace{-1mm}\orcidA{}}
\email{mainak@fnal.gov}
\affiliation{Astrophysics Theory Department, Theory Division, Fermi National Accelerator Laboratory, Batavia, Illinois 60510, USA}
\affiliation{Kavli Institute for Cosmological Physics, University of Chicago, Chicago, Illinois 60637, USA}
\author{Shunsaku Horiuchi\hspace{-1mm}\orcidB{}}
\email{horiuchi@phys.sci.isct.ac.jp}
\affiliation{Department of Physics, Institute of Science Tokyo, 2-12-1 Ookayama, Meguro-ku, Tokyo 152-8551, Japan
}
\affiliation{Kavli IPMU (WPI), UTIAS, The University of Tokyo, Kashiwa, Chiba 277-8583, Japan
}
\affiliation{Center for Neutrino Physics, Department of Physics, Virginia Tech, Blacksburg, VA 24061, USA
}
\date{\today}
\begin{abstract}
When a rapidly-rotating, highly magnetized white dwarf (WD) approaches the Chandrashekhar limit through mass accretion, it can undergo an accretion-induced collapse (AIC) to form a proto-neutron star or protomagnetar. The protomagnetar can drive a magnetically-dominated relativistic outflow, whose low entropy can lead to efficient formation of heavy nuclei. In this work, we propose that such relativistic outflows from AIC of WDs can contribute as sources of ultra high-energy cosmic rays (UHECRs). We model the acceleration of heavy nuclei in these relativistic outflows, and show that AICs can dominantly power the observed UHECRs, if a majority of them host relativistic outflows.
Accounting for uncertainties in the acceleration mechanisms and AIC rates, AICs can contribute $\sim$ a few $10^{43} - 10^{45}\ {\rm erg \ Mpc}^{-3} {\rm yr}^{-1}$ in UHECR energy generation rate density, assuming iron-like nuclei. 
\end{abstract}
\maketitle
\textit{\textbf{Introduction.}} Ultra-high energy cosmic rays (UHECRs) are the highest energy cosmic rays discovered, and their origins have remained a mystery despite decades of studies \cite{Kotera:2011cp,Anchordoqui:2018qom,AlvesBatista:2019tlv}. Only a handful of astrophysical source classes satisfy the stringent requirements set by UHECR observations. For example, the observational data demand a relatively stringent energy spectrum, a nuclear composition that increasingly becomes heavier at the highest energies, and an energy-dependent anisotropy~\cite{AlvesBatista:2019tlv}. These restrict the astrophysical source candidates to some of the most extreme objects in the Universe. Among them, powerful transients like gamma-ray bursts (GRBs) have been proposed as UHECR sources~\cite{Waxman:1995vg,Vietri:1995hs,Waxman:1997ti}.

\begin{figure}[ht!]
\includegraphics[width=0.48\textwidth]{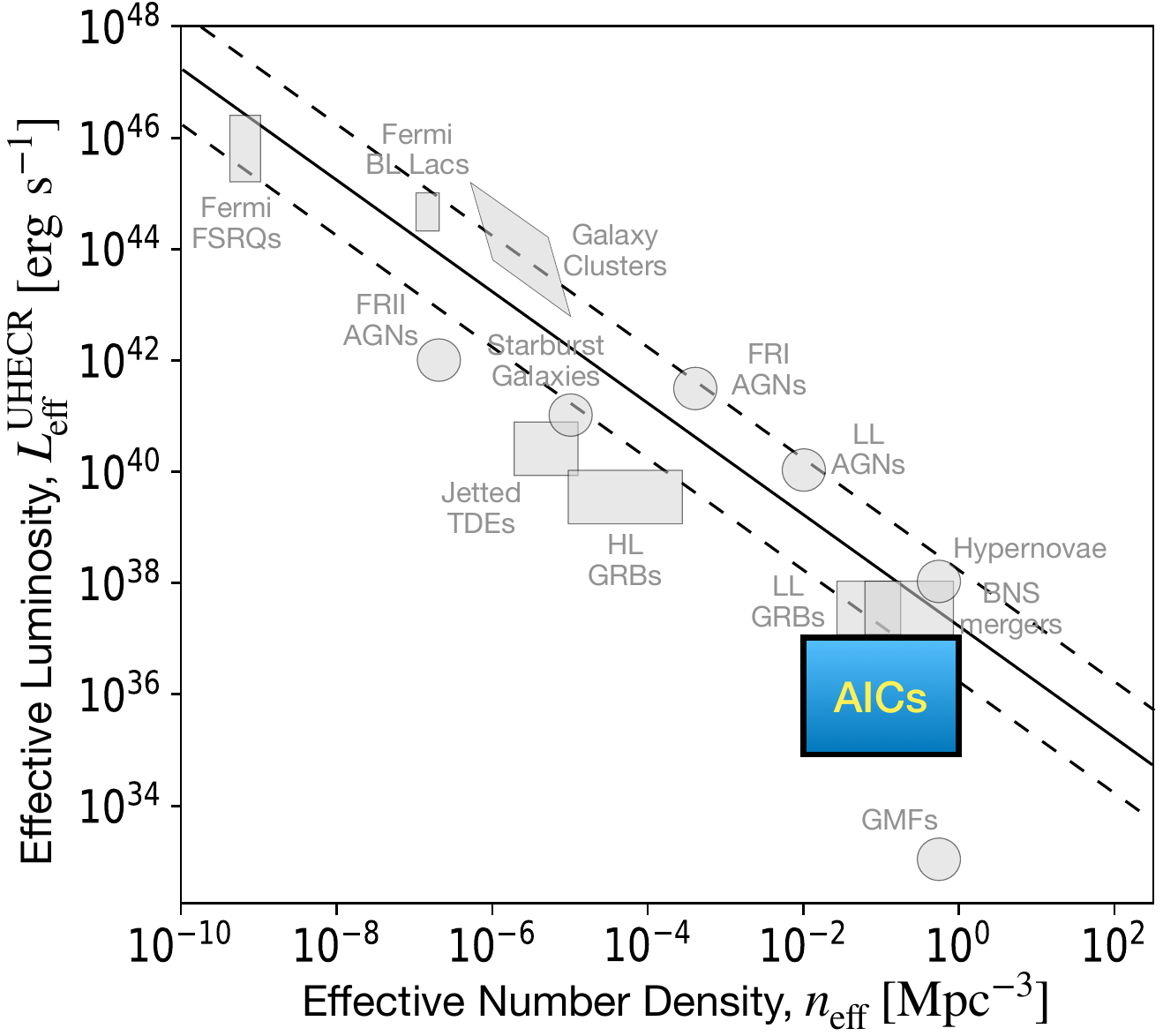}
\caption{\label{fig:result}AICs as sources of observed UHECRs on the characteristic source luminosity versus source number density plane (similar to Fig.~11 in~\cite{AlvesBatista:2019tlv}, also see references therein). The diagonal lines depict the requirements to explain the observed UHECRs, under different assumptions of the UHECR luminosity relative to photons. 
}
\end{figure}

However, the properties of the central engine powering GRBs remain an open question. While the standard paradigm invokes newly formed black hole–accretion disk systems~\cite{Woosley:1993wj,MacFadyen:1998vz,Piran:2004ba,Narayan:1992iy}, rapidly rotating magnetars provide a compelling alternative. In particular, the accretion-induced collapse (AIC) of white dwarfs has recently been proposed as a viable channel for producing such engines~\cite{Fryer:1998jb,Metzger:2007cd,Chen:2024arq,Cheong:2024hrd,Combi:2025yvs}.
WDs are the final state in the evolution of low- and intermediate-mass stars, characterized by their compact nature supported by electron degeneracy pressure. The fates of WDs depend on their intrinsic properties like mass and composition, as well as external conditions like the rate of mass accretion~\cite{NOMOTO1986249}. Most lighter mass ($\lesssim 1.2 M_\odot$) carbon-oxygen WDs encounter high accretion rates, driving a thermonuclear explosion that can result in a Type Ia supernova~\cite{Yoon:2007pw}. On the other hand, most heavier ($\gtrsim 1.2 M_\odot$) oxygen-neon dominated WDs have lower accretion rates, and can lose their electron degeneracy pressure support as a result of electron capture which drives a gravitational collapse when their mass exceeds the Chandrashekhar limit ($\sim 1.44 M_\odot$)~\cite{Chandrasekhar:1931ih}. Such collapse of a WD and the subsequent formation of a proto-neutron star is known as an accretion induced collapse (AIC)~\cite{1991ApJ...367L..19N}.

The proto-neutron star resulting from an AIC of a WD~\cite{Ablimit:2014vka,Wang:2022duk} can be rapidly rotating~\cite{Piersanti:2002sb,2003ApJ...598.1229P,Uenishi:2003sk,Saio:2004gz,2018ApJ...869..140K} and highly magnetized~\cite{2015SSRv..191..111F,2015ApJ...806L...1Z,2020AdSpR..66.1025F,Pakmor:2024efc} as a result of their mass accretion and collapse. Proto-neutron stars with millisecond spin periods and strong magnetic field strengths ($\sim {\rm a\ few\ } 10^{12} - 10^{15}$G) are regarded as protomagnetars, i.e., they will eventually settle as highly-magnetized neutron stars or magnetars. Interestingly, protomagnetars are ideal candidates to power relativistic collimated outflows (jets) using their reservoir of rotational energy, which make them attractive central engines for GRBs~\cite{Yi:1997qb,Metzger:2007cd,Perley:2008ay}. Recently, state-of-the-art general-relativistic magnetohydrodynamic (GRMHD) simulations~\cite{LongoMicchi:2023khv,Cheong:2024hrd,Combi:2025yvs} have revealed that indeed protomagnetars born as a result of AIC of WDs can launch powerful jets, supporting them as intriguing candidates for GRBs. Furthermore, simulations suggest the jets launched by protomagnetars can be neutron-rich, making them ideal sites for r-process nucleosynthesis of heavy elements~\cite{Batziou:2024ory,Yip:2024akb,Pitik:2026bjm,Pitik:2026qfj} which besides being interesting for potential kilonova observations are also relevant for the nuclear composition of potential UHECRs. 

The relativistic outflows driven by AICs are not only interesting for GRBs, but also as UHECR sources. In this \emph{Letter}, for the first time, we leverage the new connections between AICs and relativistic outflows, and in particular, recent numerical evidence suggesting that AICs can source jets powerful enough for GRBs, to propose AICs as potential UHECR sources. We consider the acceleration of heavy iron (Fe)-like nuclei in relativistic outflows launched in the AIC of WDs and their subsequent contribution to the observed UHECR emissivity. Prior work in the context of cosmic rays (CRs) from AICs~\cite{Fryer:1998jb,deGouveiaDalPino:2000jz,DeGouveiaDalPino:2001ci,1992ApJ...388..164D} has focused on CR protons accelerated just above the magnetosphere, where the protons are sourced from the neutron star crust. In contrast, we focus on a qualitatively new regime involving the acceleration of heavy-nuclei, motivated by recent simulation developments. 

Our main results are shown in Fig.~\ref{fig:result}, which shows the extent to which AIC of WDs may help explain the observed UHECRs. We conclude that despite large uncertainties, a population of AICs can indeed serve as potential UHECR sources at the current observational level. Furthermore, this is particularly timely given AICs have already been proposed to explain kilonova observations from GRB 21211A~\cite{Mei:2022ncd,Rastinejad:2022zbg,Troja:2022yya,Yang:2022qmy} and GRB 230307A~\cite{2023GCN.33429....1B,Dalessi:2025unu,JWST:2023jqa} and are predicted to be observed in the Vera C.~Rubin Observatory Legacy Survey of Space and Time (LSST), the James Webb Space Telescope (JWST), and in MeV gamma-rays (using COSI~\cite{Tomsick:2023aue} and AMEGO-X~\cite{Caputo:2022xpx}). Our work not only  highlights the further importance of AICs as a UHECR source, but also opens future studies to potentially test this scenario.

\textit{\textbf{Modeling of relativistic outflows in AICs.}}
\begin{figure}
\includegraphics[width=0.48\textwidth]{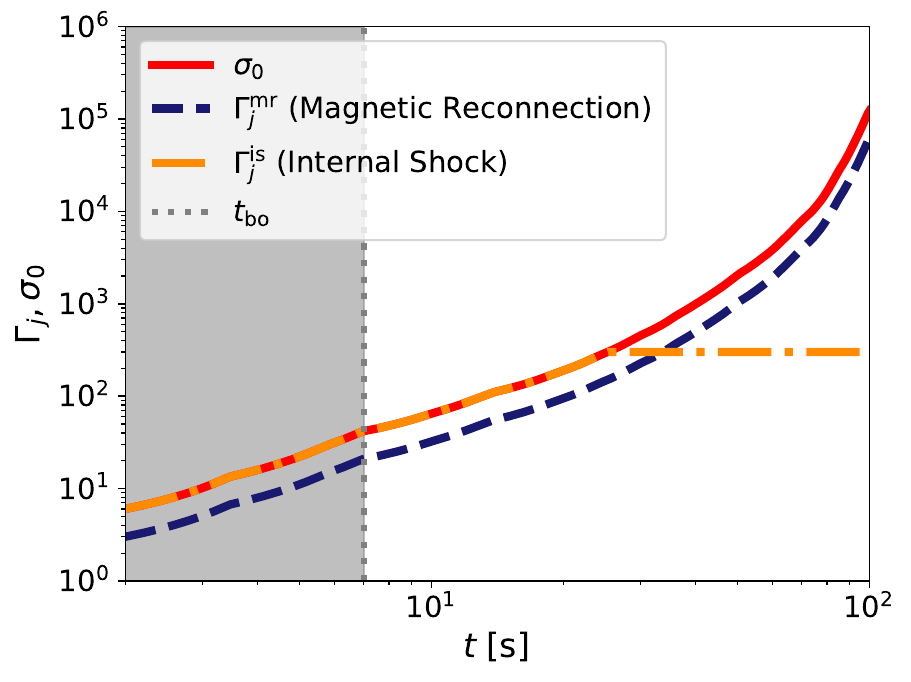}
\caption{\label{fig:jet_mod}Time evolution of the magnetization $\sigma_0$ (Eq.~\ref{eq:sigma0}) and the terminal bulk Lorentz factor $\Gamma_j$ of the jet for the magnetic reconnection and internal shock scenarios. The breakout time ($t_{\rm bo}$) prior to which $\Gamma_j$ may not reach the terminal value is shaded in gray.
}
\end{figure}
After a brief mass accreting phase, a newly-born protomagnetar will spin down due to loss of its rotational energy through magnetically-driven pulsar winds, on a timescale known as the spindown timescale estimated by $t_{\rm sd} \approx E_{\rm rot}^{\rm init}/L^{\rm init}_{\rm sd}$. 
Here, $E_{\rm rot}^{\rm init} \approx (1/2) I \Omega_i^2$ is the initial rotational energy, $I = (4/10) M_* R_*^2$ is the moment of inertia with mass $M_*$ and radius $R_*$ of the pulsar, $\Omega_i = 2\pi/P_i$ is the initial angular velocity, and $P_i$ is the initial spin period. The dipolar magnetic field dominates over the toroidal fields near the protomagnetar surface and is given by $B_d$ such that the magnetic dipole moment $\mu = B_d R_*^3$. The initial spin down luminosity is then $L^{\rm init}_{\rm sd} = (\chi/c^3) \mu^2 \Omega_i^4$, where $\chi$ takes into account the inclination between the rotation and magnetic axes, $\chi \sim \sin \iota$, where $\iota$ is the inclination angle.

In estimating the properties of the outflow sourced by the protomagnetar in the AIC of a WD, we follow an approach similar to Refs.~\cite{Metzger:2010pp,Metzger:2011xs}. Its magnetization or equivalently baryon loading is estimated as
\be
\label{eq:sigma0}
\sigma_0 (t) = \frac{\phi^2 \Omega(t)^2}{\dot{M}(t) c^3}\,,
\ee
where the open magnetic flux of the protomagnetar (assuming it to be a rotating dipole) $\phi \approx f_{\rm op} B_d R_*^2$, the angular velocity $\Omega (t) = 2\pi/P(t)$, $f_{\rm op}$ is the fraction of the magnetar surface threaded by open field lines, the spin period $P(t) = P_i \big( 1 + t/t_{\rm sd}\big)^{1/2}$, and $\dot{M}(t)$ is the mass loss rate for a strongly magnetized outflow defined as
\be
\label{eq:masslossrate}
\dot{M}(t) = \begin{cases}
\dot{M}_\nu f_{\rm op}\,,\hspace{4em}(\iota + \theta_{\rm op}) \ll \pi/2\\
\dot{M}_\nu f_{\rm op} f_{\rm cent}\,,\hspace{2em}(\iota + \theta_{\rm op}) \gtrsim \pi/2\\
\dot{M}_{\rm GJ}\,,\hspace{5em}\dot{M} \leq \dot{M}_{\rm GJ}
\end{cases}\,,
\ee
where $\dot{M}_\nu$ is the mass loss rate caused by neutrino-heating \cite{Qian:1996xt}, $f_{\rm cent}$ is the enhancement to the mass loss rate as a result of changes in the effective potential due to centrifugal forces, and $\theta_{\rm op}$ is the half-opening angle of the polar cap, i.e., it quantifies the extent of the region beyond which the magnetic field lines open up and form the wind region, that is, for an angle $\theta > \theta_{\rm op}$ the magnetosphere is closed, while it is open for $\theta<\theta_{\rm op}$. The Goldreich-Julian mass loss rate given by $\dot{M}_{\rm GJ}$, is only relevant at very late times when the protomagnetar becomes transparent to neutrinos and has no impact on our results. The relevant details for computing the mass loss rate are provided in Appendix~\ref{appsec:massloss_rate}.

The protomagnetar is a copious source of neutrinos, but this rapidly declines in time. This drives down $\dot{M}$, causing the outflow magnetization $\sigma_0$ to rapidly increase, as shown in Fig.~\ref{fig:jet_mod} (red solid). 
After launch, the outflow is mainly accelerated by magneto-centrifugal forces, i.e., it converts a significant fraction of its magnetic energy into bulk kinetic energy to attain the high ($>100$) bulk Lorentz factors needed for GRBs. The magnetization $\sigma_0$ determines the final terminal bulk Lorentz factor of the outflow. The bulk energy of the relativistic outflow can power a GRB only after it breaks out of the expanding supernova ejecta. For typical GRB progenitors, the jet breakout time is $\mathcal{O}(10)$ s, implying the jet's bulk Lorentz factor is $\sigma_0 \gtrsim 100$ (see Fig.~\ref{fig:jet_mod}). Additionally, the GRB needs to happen directly near or above the photosphere; otherwise, high-energy emission will be partially thermalized and suppressed. The prompt emission phase, if a GRB is produced, is assumed to terminate when the magnetization becomes too large for efficient dissipation, $\sigma_0 \sim 10^4$ over the next tens of seconds~\cite{Zhang:2003uk,Meszaros:2006rc}. 

We adopt $M_* = 1.42 M_\odot$, $R_* = 10$ km, $P_i = 2$ ms, $B_d = 5 \times 10^{15}$ G, and $\chi = 1$ ($\iota = \pi/2$) as our canonical values, such that $L^{\rm init}_{\rm sd} = 9 \times 10^{49}$ erg/s and $t_{\rm sd} = 6.18 \times 10^1$ s. The jet (collimated relativistic part of the outflow) luminosity $L_j(t) = L_{\rm sd}^{\rm init} \big( 1 + t/t_{\rm sd} \big)^{-2}$, from which the isotropic equivalent jet luminosity is $L_{j,\rm iso} = L_j/f_{\rm bm}$, where the beaming factor $f_{\rm bm} = (1 - \cos\ \theta_j) \sim (1/2)\theta_j^2$ assuming that the half-opening angle of the jet, $\theta_j \simeq 8^\circ$ is small. The isotropic equivalent jet energy is given by $E_{j,\rm iso} = \int dt\ L_{j,\rm iso} (t)$. For our adopted parameters, $E_{j,\rm iso} = 3.7 \times 10^{53}$ erg while $E_j = f_{\rm bm} E_{j,\rm iso} = 3.6 \times 10^{51}$ erg. This is larger than the early-time isotropic-equivalent jet energy seen in current GRMHD simulations, e.g., $E_{j,\rm iso} \sim 8 \times 10^{51}$ erg at $t \sim 1$ s in Ref.~\cite{Cheong:2024hrd}, but those calculations are not evolved over the full burst duration and clearly show that the jet energy is still increasing (also see~\cite{Combi:2025yvs}). Furthermore, the breakout time of the jet is given by $t_{\rm bo} \sim 7\ {\rm s}\ (R_{\rm surf}/10^{11}\ {\rm cm}) (\beta/0.5)^{-1}$, where $R_{\rm surf}$ is the distance by which the jet propagates before breaking out and $\beta c$ is its velocity during propagation within the stellar envelope. To ensure that the jet reaches the terminal value of the Lorentz factor, we only focus on the regime when $t>t_{\rm bo}$.

\textit{\textbf{Acceleration of CRs.}}
\begin{figure*}
\includegraphics[width=0.9\textwidth]{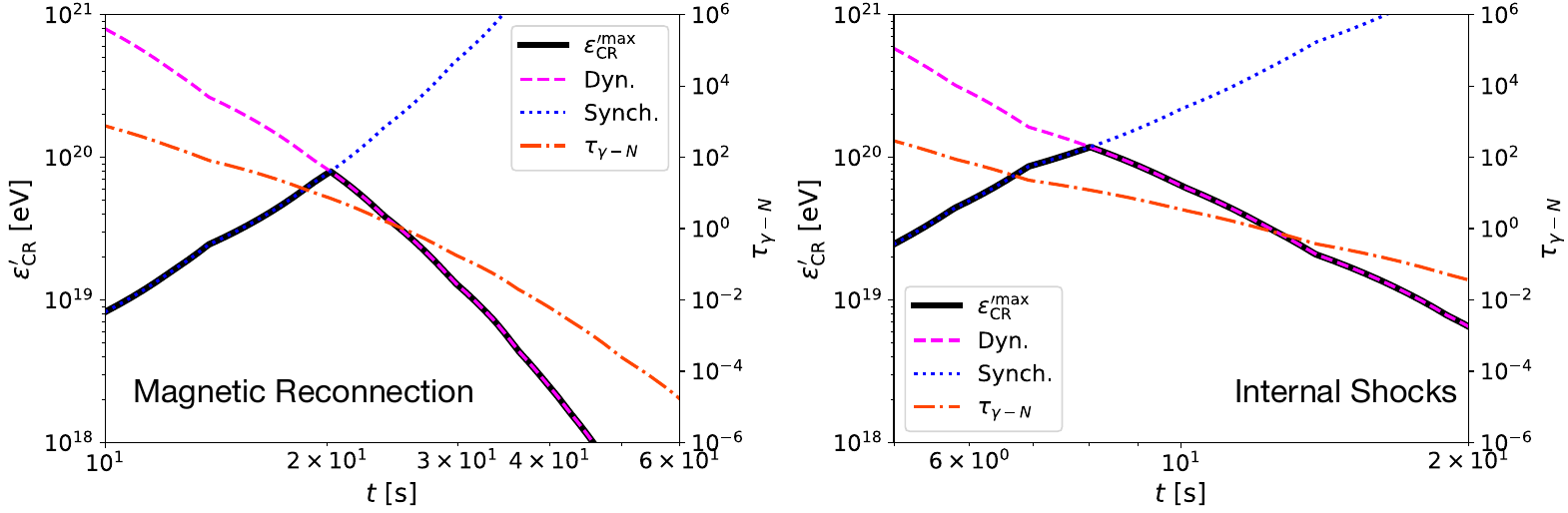}
\caption{\label{fig:magreconn_intshk}Time evolution of the comoving maximum CR energy $\varepsilon_{\rm CR}^{\prime \rm max}$ that can be achieved in the magnetic reconnection \emph{(left)} and internal shock \emph{(right)} scenarios. The number of photon-nuclei interactions are also shown on the right axes for each figure, where we assume the nuclei survives for $\tau_{\gamma - N} \leq 10$.
}
\end{figure*}
If the relativistic jet powers a GRB, the prompt emission is expected to arise when the jet dissipates its energy through processes such as magnetic reconnection and/or internal shocks. Motivated by this, we consider UHECR acceleration at analogous sites, through the same processes. Following Ref.~\cite{Metzger:2011xs}, we assume that the acceleration of charged nuclei occurs at the same location as the dissipation region responsible for the GRB emission. We denote quantities in the jet comoving (jet rest) frame with a prime, e.g., the energy of a CR species is  $\varepsilon_{\rm CR}^\prime$ and its energy in the lab frame is $\varepsilon_{\rm CR} \approx \Gamma_j \varepsilon_{\rm CR}^\prime$. The maximum energy $\varepsilon_{\rm CR}^{\prime \rm max}$ to which CRs can be accelerated in the jet can be obtained by balancing the acceleration timescale $t_{\rm acc}$ with various cooling timescales $t_{\rm cool}$ and the dynamic timescale $t_{\rm dyn}$. In other words, the maximum CR energy is obtained by solving for $t_{\rm acc}^\prime(t,r,\varepsilon_{\rm CR}^{\prime \rm max}) = t_{\rm cool}^\prime(t,r,\varepsilon_{\rm CR}^{\prime \rm max})$ and $t_{\rm acc}^\prime(t,r,\varepsilon_{\rm CR}^{\prime \rm max}) = t_{\rm dyn}^\prime(t,r)$. Below, we consider two scenarios -- magnetic reconnection within the jet, and internal shocks formed when faster parts of the jet catch up and collide with slower parts of the jet.

\textit{Magnetic Reconnection.}
For rapid magnetic reconnection, the corresponding acceleration timescale is determined by the Larmor radius ($r_L$) and can be estimated
\be
t_{\rm acc}^\prime (t,r = R_{\rm mr},\varepsilon_{\rm CR}^\prime) \sim \eta_{\rm acc}\frac{r_L}{c} \approx \eta_{\rm acc} \bigg( \frac{2 \pi \varepsilon_{\rm CR}^\prime}{(Z e) B^\prime (t,R_{\rm mr}) c} \bigg)\,,
\ee
where $\eta_{\rm acc}$ is the acceleration efficiency that includes the micro-physical details such as the reconnection rate, particle scattering in turbulent magnetic fields, and losses and escape from the reconnection region~\cite{Lyubarsky:2005zt,2009ApJ...700L..16D,Sironi:2014jfa}. We parameterize all of this by $\eta_{\rm acc} (= 1)$. The saturation radius $R_{\rm mr}$ is the radius beyond which the Lorentz factor becomes constant, that is, the flow attains its terminal velocity. This is given by, $R_{\rm mr}(t) \approx \sigma_0^2(t) P(t) c/ (6 \beta_{\rm mr})$, where $\beta_{\rm mr} = v_{\rm mr}/c$ gives the reconnection speed $v_{\rm mr}$ normalized to $c$ and we adopt $\beta_{\rm mr} = 0.01$, ignoring any radial dependence. Typically, $R_{\rm mr}\sim 10^{13}$ cm at relevant times; we show the overall evolution in Appendix~\ref{appsec:reg_partacc}. The comoving magnetic field strength is derived imposing $B^{\prime 2} (t,r)/(8\pi) = \epsilon_B u^{\prime}(t,r)$, where the $u^{\prime}(t,r)$ is the comoving isotropic equivalent energy density and $\epsilon_B$ is the fraction of energy dissipated to magnetic fields. This gives,
\be
\label{eq:bprime}
B^\prime (t,r) \simeq \sqrt{2 \epsilon_B L_{j,\rm iso}(t) r^{-2} \Gamma_j^{-2}(t) c^{-1}}\,.
\ee
For magnetic reconnection the Lorentz factor at the acceleration site, $\Gamma_j(t) = \Gamma_j^{\rm mr}(t) \sim \sigma_0(t)/2$, which increases with time since the jet's $\sigma_0$ increases (Fig.~\ref{fig:jet_mod}). The atomic number and mass number of the CR species is denoted by $Z$ and $A$ respectively, $c$ is the speed of light, $m_p$ is the mass of proton, and $e$ is the electric charge, which implies $A m_p$ and $Z e$ give the total mass and charge of a given CR species. We focus on iron (Fe) in this work as a representative case, i.e., we set $A = 56$ and $Z = 26$. For magnetic reconnection, we choose $\epsilon_B = 1$ and $r = R_{\rm mr}$. 

\textit{Internal Shocks.}
Since the outflow launched later in time obtains larger terminal Lorentz factors, it naturally catches up with earlier parts of the outflow. Thus, the outflow evolution is conducive to collisions, and therefore, shocks; we call this the internal shock scenario (see Refs.~\cite{Metzger:2010pp,Metzger:2011xs}). We require internal shocks to occur outside the photosphere radius~\cite{Meszaros:2006rc},
\be
R_{\rm ph}(t) \approx \frac{L_{k,\rm iso}(t) \sigma_T}{8 \pi m_p c^3 \Gamma_j^3(t)} \approx \frac{L_{j,\rm iso}(t) \sigma_T}{(1+\sigma_0) 8 \pi m_p c^3 \Gamma_j^3(t)}\,,
\ee
where $L_{k,\rm iso}$ is the isotropic-equivalent kinetic luminosity of the jet, $\sigma_T$ is the Thomson cross-section, and $\sigma_0$ sets the initial baryon loading. We define the magnetization at the internal shock $\sigma = L_B/L_k$\footnote{Note that this is different from $\sigma_0$ which describes the magnetization at launch. After launch the jet is accelerated efficiently, i.e., a substantial $L_B$ will be converted to $L_k$, and the residual local magnetization is given by $\sigma$.} where $L_k + L_B = L_j$. Thus, $L_{k,\rm iso} = L_{\rm j,iso}/(1+\sigma)$. The magnetization at the internal shock is $\sigma(r = R_{\rm is}) \ll 1$. The internal shock radius $R_{\rm is}$ can be estimated using $R_{\rm is}(t) \approx 2 \Gamma_j^2(t) c t_j$, where $t_j$ denotes the time at which the faster moving shell was launched with respect to the bulk \cite{Metzger:2010pp}.
Thus, larger value of $t_j$ lead to larger values of $R_{\rm is}$. Demanding the shock to not be radiation mediated thus sets
\be
t_j(t) \gtrsim \frac{R_{\rm ph}(t)}{2 \Gamma_j^2 (t) c} \approx \frac{L_{j,\rm iso}(t)\sigma_T}{(1+\sigma_0) 16 \pi m_p c^4 \Gamma_j^5 (t)}\,.
\ee
For the relevant CR-acceleration timescales, we choose a typical value for $t_j = 1.0$ s. The variation of $R_{\rm is}$ and $R_{\rm ph}$ for different values of $t_j$ is discussed in Appendix~\ref{appsec:reg_partacc}. We assume that the bulk Lorentz factor $\Gamma_j^{\rm is} \approx {\rm min} \left[ \sigma_0, \bar{\Gamma}_{j}^{\rm is} \right]$, where we assume $\bar{\Gamma}_{j}^{\rm is} = 300$ is the typical GRB Lorentz factor when the flow saturates~\citep{Waxman:1995vg,Murase:2008mr}, that is, two shells emitted with different Lorentz factors merge and move with a common effective Lorentz factor $\bar{\Gamma}^{\rm is}$. Similar to the magnetic reconnection case, here we evaluate the comoving magnetic field strength $B^\prime$ using Eq.~\eqref{eq:bprime}, with $\epsilon_B = 0.1$ but with $r = R_{\rm is}$.

The results for the maximum CR energy achievable are shown in Fig.~\ref{fig:magreconn_intshk} for magnetic reconnection \emph{(left)} and internal shock scenarios \emph{(right)}, taking into account energy losses (discussed below). We see that $\varepsilon_{\rm CR}^{\prime \rm max} \sim 8.0 \times 10^{19}$ eV can be achieved for magnetic reconnection while the same for the internal shock scenario is $\varepsilon_{\rm CR}^{\prime \rm max} \sim 1.2 \times 10^{20}$. The CR spectra can be estimated using a power law with an exponential decay for $\varepsilon_{\rm CR}^\prime > \varepsilon_{\rm CR}^{\prime \rm max}$,
\be
\label{eq:crspectra}
\frac{dN}{d \varepsilon_{\rm CR}^\prime}  = \mathcal{N} \varepsilon_{\rm CR}^{\prime -p}\  {\rm exp}\bigg(-\frac{\varepsilon_{\rm CR}^\prime}{\varepsilon_{\rm CR}^{\prime \rm max}} \bigg)\,,
\ee
where $\mathcal{N}$ is the normalization and $p$ indicates the spectral index. We choose $p = 2$, but discuss the uncertainty due to other values in Appendix~\ref{appsec:uncertainty}. The normalization is evaluated by assuming a fraction $\eta_{\rm CR}$ of the true jet energy goes into CRs such that
\be
\mathcal{N} = \frac{\eta_{\rm CR} E_j}{\int_{\varepsilon_{\rm CR}^{\rm low}}^{\varepsilon_{\rm CR}^{\rm up}}\ d\varepsilon_{\rm CR}^\prime\  \varepsilon_{\rm CR}^{\prime} \varepsilon_{\rm CR}^{\prime -p}\  {\rm exp}\big(-\varepsilon_{\rm CR}^\prime/\varepsilon_{\rm CR}^{\prime \rm max}\big)}\,,
\ee
where $\varepsilon_{\rm CR}^{\rm low} = 10^{18}$ eV and $\varepsilon_{\rm CR}^{\rm up} = 10^{21}$ eV are the lower and upper integration limits. We assume a commonly adopted $\eta_{\rm CR} = 0.1$, \emph{i.e.}, $10\%$ of the jet energy goes into UHECRs. In Appendix.~\ref{appsec:reg_partacc}, we further show that despite the baryon-poor nature of the outflow, the required number of accelerated Fe-like UHECRs constitutes only a small fraction of the available heavy-nuclei content.

\textit{\textbf{Energy losses.}}
The cooling timescale is dominated mostly by synchrotron cooling \cite{Metzger:2011xs} and therefore we ignore other cooling channels including inverse Compton, diffusion, Bethe-Heitler, photohadronic, and hadronuclear losses. However, we do take into account photodisintegration by photons. 

\textit{Synchrotron and dynamical losses.} The synchrotron timescale $t_{\rm sync}$ can be defined as
\ba
\label{eq:tsync}
t_{\rm sync}^\prime(t, r, \varepsilon_{\rm CR}^\prime) &\approx& 3 \bigg(\frac{A m_p c}{Z e}\bigg)^4 \bigg(\frac{c^3 \Gamma_j(t)}{B^{\prime 2}(t,r) \varepsilon_{\rm CR}^\prime}\bigg) \nonumber \\
&=& 3 \bigg(\frac{A m_p c^2}{Z e}\bigg)^4 \bigg(\frac{\Gamma_j^3(t) r^2}{2 \epsilon_B L_{j,\rm iso}(t) \varepsilon_{\rm CR}^\prime}\bigg)\,,
\ea
where Eq.~\eqref{eq:bprime} defines the comoving magnetic field strength. The comoving dynamical timescale is given by the expansion of the system, such that
\be
\label{eq:tdyn}
t_{\rm dyn}^\prime(t,r) \approx r/\big( \Gamma_j(t) c \big)\,.
\ee
The synchrotron cooling timescale and the dynamical timescale for the magnetic reconnection scenario with $r = R_{\rm mr},\ \Gamma_j = \Gamma_j^{\rm mr}$ and the internal shock scenario with $r = R_{\rm is},\ \Gamma_j = \Gamma_j^{\rm is}$, are estimated using Eqs.~\eqref{eq:tsync} and~\eqref{eq:tdyn} and shown in Fig.~\ref{fig:magreconn_intshk} for the respective cases.

\textit{Photodisintegration.}
To contribute to the observed flux of nuclei UHECR, the Fe nuclei besides getting accelerated to ultra high-energies also need to avoid complete photodisintegration at the source and also while propagation from the source to Earth. Since we consider a pure Fe composition at source, the giant dipole resonance (GDR) dominates over pion production and other loss mechanisms in the context of photodisintegration. We consider GRB photons with characteristic energy $E_{\rm pk} \sim 0.1 - 1$ MeV as the source of photons for photodisintegration. The number of interactions experienced by an UHECR can be estimated
\be
\label{eq:photodisint}
\tau_{\gamma-N} (t,r) \approx \frac{L_{j,\rm iso}(t) \epsilon_{\rm rad} \mathcal{C} \sigma_{\rm GDR}  (\Delta \epsilon_{\rm GDR}/\bar{\epsilon}_{\rm GDR})}{4 \pi E_{\rm pk}  r c \Gamma_j(t)^2}\,,
\ee
where $\epsilon_{\rm rad}$ is the radiative efficiency of the jet, $\mathcal{C}$ is the fraction of gamma-ray energy released below the band peak $E_{\rm pk}$ (see~\cite{Metzger:2011xs} for more details), $\sigma_{\rm GDR}$, $\Delta \epsilon_{\rm GDR}$, and $\bar{\epsilon}_{\rm GDR}$ denotes the GDR cross-section, energy, and width respectively.

We choose $\epsilon_{\rm rad} = 0.1$, $\mathcal{C} = 0.1$, $\sigma_{\rm GDR} = 8 \times 10^{-26} (A/56)\ {\rm cm}^{2}$, the typical energy for GDR is given by $\bar{\epsilon}_{\rm GDR} \sim 18\ {\rm MeV} A_{56}^{-0.21}$, the line-width $\Delta \epsilon_{\rm GDR}/\bar{\epsilon}_{\rm GDR} \sim 0.4 (A/56)^{0.21}$, and $E_{\rm pk} = 1$ MeV. The results for photodisintegration at the source, for each acceleration mechanism, are shown in Fig.~\ref{fig:magreconn_intshk}. For the survival criterion we will adopt $\tau_{\gamma-N} < 10$, that is, a nucleus can undergo approximately $10$ interactions before its composition is significantly altered. Thus in both acceleration mechanisms, the composition of UHECR can remain heavy at the source.

Photodisintegration is also important during propagation, if these UHECRs are to reach Earth. Here again, the GDR interactions with extragalactic infrared background (EIB) provide the dominant process and target. This is because the photon energy in the lab (or EIB) frame is given by $\epsilon_{\gamma}^{\rm EIB} = \bar{\epsilon}_{\rm GDR}/\gamma_A$, where $\gamma_A = E_A/(A m_p c^2)$ is the Lorentz factor of the nuclei. Thus, $\epsilon_{\gamma}^{\rm EIB} \sim 10^{-3}\ {\rm eV} A_{56} E_{20}^{-1} > \epsilon_\gamma^{\rm CMB} \approx 6 \times 10^{-4}$ eV~\cite{Fixsen:2009ug}. Assuming the EIB spectrum to be $\propto (E_\gamma^{\rm EIB})^{-2.5}$, we have the mean free path $\lambda_{\rm mf} \sim 1/\big( N_\gamma(E = \epsilon_{\gamma}^{\rm EIB}) \times \sigma_{\rm GDR} \times (\Delta \epsilon_{\rm GDR}/\bar{\epsilon}_{\rm GDR})\big) \approx 10\ {\rm Mpc}\ E_{20}^{-1.5} A_{56}^{0.3}$, by normalizing the EIB photon number density as in Ref.~\cite{Allard:2005ha,Metzger:2011xs}. Therefore, $\lambda_{\rm mf}$ provides the length before the UHECR nuclei suffer one interaction. 

For multiple interactions, we estimate the length at which the nuclei lose $25\%$ of their initial energy ($\chi_{75}$) by taking into account the change in $\lambda_{\rm mf}$ as the nuclei are changed as a result of successive photodisintegrations. We assume that the mean number of nucleons ejected per disintegration $\bar{n}_{\rm ej} \approx 1$, which in the lab-frame have the same Lorentz factor as the parent nuclei\footnote{This is because the nucleons are ejected at sub-relativistic velocities, so their relative Lorentz factor with respect to the parent nuclei can be neglected.}. The decrease in energy for the parent nuclei $E_A \propto A$ and $\gamma_A \sim$ constant, which implies that the target EIB energy and hence $N_\gamma^{\rm EIB}$ that the nuclei interacts with while propagating is the same for the entire propagation. Moreover, from above $ \sigma_{\rm GDR} \times (\Delta \epsilon_{\rm GDR}/\bar{\epsilon}_{\rm GDR}) \propto A^{1.2}$ and hence $\lambda_{\rm mf} \propto A^{-1.2}$. Thus, we finally have $\chi_{75} \approx \bar{n}_{\rm ej}^{-1} \sum_{i = 3A/4}^A \lambda_{{\rm mf},i} \simeq 170\ {\rm Mpc}\ E_{20}^{-1.5} A_{56}^{1.3} \bar{n}_{\rm ej}^{-1}$. Since we are mostly interested in nearby AICs to contribute to the UHECR flux, our estimates show that indeed for typical distances of $\mathcal{O}(100\ {\rm Mpc})$, the UHECRs survive the propagation to the Earth. We note that this is a very rough estimate and a detailed propagation calculation can concretely explore the propagation effects, which we defer to future work.

\textit{\textbf{Rate of AIC\MakeLowercase{s}}.}
The true $z=0$ rate density of AICs can be approximated by,
\be
\label{eq:aicrate}
\dot{\rho}_{\rm AIC}^{\rm true} \approx n_g^{\rm MW} R_{\rm AIC}^{\rm MW}\,,
\ee
where $n_g^{\rm MW}$ is the number density of Milky Way (MW) type galaxies and $R_{\rm AIC}^{\rm MW}$ is the true rate of AICs in a MW type galaxy. The typical number density of MW type galaxies is $n_g^{\rm MW} \sim 10^{-2}\ {\rm Mpc}^{-3}$~\cite{2016ApJ...830...83C}, while the true rate of AICs in MW type galaxy is uncertain but $R_{\rm AIC}^{\rm MW} \sim 10^{-6} - 10^{-4}\ {\rm yr}^{-1}$ from population synthesis models~\cite{Fryer:1998jb,Dessart:2006gd,Ruiter:2018ouw}. This results in $\dot{\rho}_{\rm AIC}^{\rm true} \sim 10^{-8} - 10^{-6}\ {\rm Mpc}^{-3}{\rm yr}^{-1}$. We choose $\dot{\rho}_{\rm AIC}^{\rm true} = 10^{-7}\ {\rm Mpc}^{-3}{\rm yr}^{-1} (= 100\ {\rm Gpc}^{-3}{\rm yr}^{-1})$ as our fiducial value. Since we are limited to nearby $\mathcal{O}(100\ {\rm Mpc})$ sources for contributing to the observed UHECR, redshift evolution can be ignored. The apparent rate ($\dot{\rho}_{\rm AIC}^{\rm app}$) can be computed using $\dot{\rho}_{\rm AIC}^{\rm app} = f_{\rm bm} \dot{\rho}_{\rm AIC}^{\rm true}$. The fraction of AICs with relativistic outflows is highly uncertain. One metric is the GRB rate; by assuming that the relativistic outflows in AICs power GRBs, the GRB rate may be adopted instead, which is approximately a tenth of the AIC rate we adopt. However, whether relativistic outflows from AICs always power long GRBs is unclear and will depend on details of the outflow properties, GRB models, and whether related transients such as low-luminosity GRBs and X-ray flashes are considered. We therefore leave such a population analysis for future work; our adopted AIC rate should therefore be taken as an upper limit for the maximum number of AICs with relativistic outflows.

\textit{\textbf{Results and discussions.}}
\begin{figure}
\includegraphics[width=0.48\textwidth]{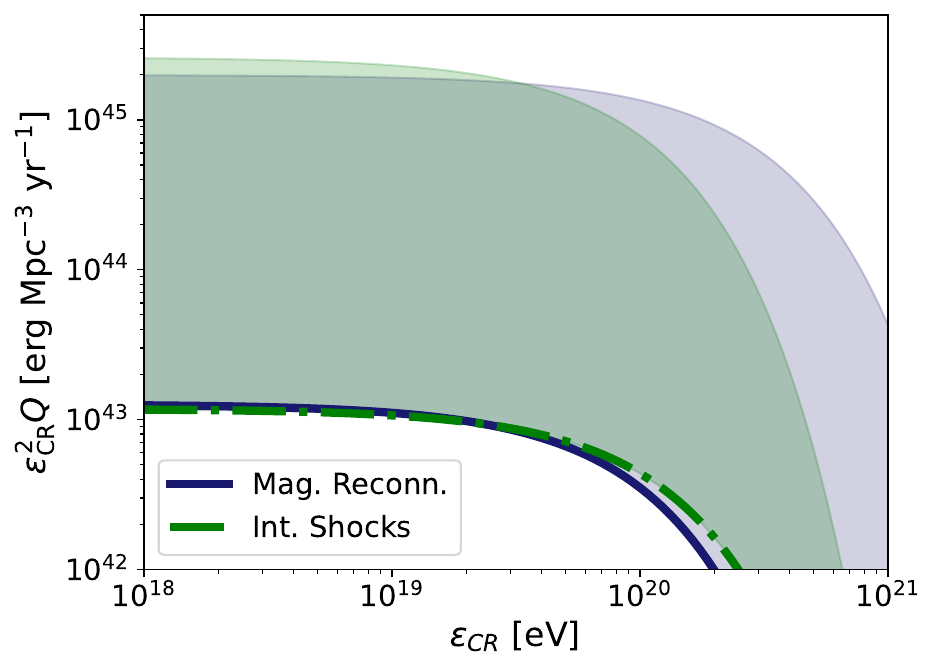}
\caption{\label{fig:res_1}The contribution of jets originating from AIC of WDs to the UHECR emissivity for pure Fe-like composition for magnetic reconnection and internal shock scenarios (Eq.~\ref{eq:contri}). The solid lines show our estimate with fiducial parameters discussed in the text while the shaded band encapsulates the various uncertainties to show plausible upper limits (see Appendix~\ref{appsec:uncertainty}). 
}
\end{figure}
To evaluate the contribution of AICs to the UHECR emissivity for Fe-like composition, we use the same phenomenological spectral form as in Eq.~\eqref{eq:crspectra} for the source-frame spectrum. Further, we conservatively set $\varepsilon_{\rm CR} \approx \varepsilon_{\rm CR}^\prime$, i.e., we do not introduce an additional Lorentz boost from the jet Lorentz factor when estimating the lab frame energy, and we obtain,
\be
\label{eq:contri}
\varepsilon_{\rm CR}^2 Q(\varepsilon_{\rm CR}) = \dot{\rho}_{\rm AIC}^{\rm true} \bigg(\varepsilon_{\rm CR}^2 \frac{dN}{d\varepsilon_{\rm CR}} \bigg)\,,
\ee
which can be evaluated by using Eq.~\eqref{eq:aicrate}, with the corresponding $\varepsilon_{\rm CR}^{\prime \rm max}$ obtained from the acceleration mechanism used. The results for the fiducial case are shown as thick lines in Fig.~\ref{fig:res_1}, while the plausible results due to various parameters are shown as a band and discussed in Appendix~\ref{appsec:uncertainty}. Including the Lorentz boost would increase the observed maximum energy $\varepsilon_{\rm CR}$. However, since we have fixed $\bar{\Gamma}_{j}^{\rm is} = 300$, this boost is arbitrary even though it is motivated. Furthermore, it has little bearing on the observed UHECRs, since the spectrum above the $\sim 10^{20}$ eV will be suppressed due to interactions with diffuse photon backgrounds (the GZK cutoff). The interpretation of the observed UHECR cutoff remains uncertain. We note in this context that a detailed calculation of our model, taking into account time dependence of the UHECR source flux normalization and Lorentz boost (hence maximum energy) will be needed to provide an interpretation of whether the observed cutoff is due to source acceleration physics or propagation effects. We leave this to future work.

To quantify the contribution of AIC of WDs to the observed UHECR, we consider the source number density and luminosity. The effective number density of AIC of WDs is $n^{\rm AIC}_{\rm eff} = (3/5)\rho_{\rm AIC}^{\rm app} \tau_{\rm AIC}$, where $\tau_{\rm AIC}$ is the apparent burst duration of the UHECR burst and can be approximated by $\tau_{\rm AIC} = d_L^2 Z^2 \langle B_{\rm mag}^2 \lambda_{\rm corr}\rangle /\big( 9 E^2 \big)$, where $d_L$ is the typical distance traveled by the UHECR from the source, $\lambda_{\rm corr}$ is the correlation length of the magnetic field, and $B_{\rm mag}$ is the magnetic field strength encountered by the UHECR while propagating. Furthermore, the effective luminosity of the UHECRs from AIC of WDs is $L_{\rm eff}^{\rm UHECR, AIC} \approx \big(\int d\varepsilon_{\rm CR}\ \varepsilon_{\rm CR} dN/d\varepsilon_{\rm CR}\big)/\tau_{\rm AIC}$. We assume typical values of $B_{\rm mag} \sim 1$ nG, $\lambda_{\rm corr} \sim 1$ and $d_L \sim 100$ Mpc. In Fig.~\ref{fig:result}, we show where AICs would lie in the $L_{\rm eff}^{\rm UHECR, AIC} - n^{\rm AIC}_{\rm eff}$ plane along with a variety of other sources (see~ Refs.~\cite{AlvesBatista:2019tlv,Murase:2008sa} for more details). The diagonal lines represent the required values to power the observed UHECRs, under the assumption that the UHECR luminosity is $1, 0.1, 10.0$ times their photon luminosities.

Remarkably, we note that indeed given the large uncertainty in the rate of AIC of WDs and their energy budget, they have sufficient capability to power UHECRs to contribute to the observed UHECR flux. In the era of Rubin LSST, JWST, and other powerful transient telescopes, the current uncertainties will decrease which will then help in testing our hypothesis presented in this work. In fact, a more detailed calculation encompassing the various effects at the microphysics level regarding the maximum energy achievable by the CRs in the jet, time dependence, a population of heavy nuclei that includes other r-process elements, their survival, and propagation effects will need to be performed in the future. Our current work serves as a step in that direction and highlights the potential importance of AIC of WDs as UHECR sources.

\acknowledgements
\textit{\textbf{Acknowledgements.}} We wish to thank the organizers of WHEPP 2025 where this idea originated. We thank Damiano Francesco Giuseppe Fiorillo, David Radice, and Roland Crocker for comments and careful reading of our manuscript. M.\,M. is grateful to the Centre for Advanced Study at the Norwegian Academy of Science and Letters for their hospitality and support where the last stages of this work was completed. M.\,M. acknowledges support from the FermiForward Discovery Group, LLC under Contract No. 89243024CSC000002 with the U.S. Department of Energy, Office of Science, Office of High Energy Physics. SH acknowledges NSF Grant No.~PHY-2209420 and JSPS KAKENHI grant No.~23H04899. 
\bibstyle{aps}
\bibliography{refs}

\begin{thebibliography}{72}%
\makeatletter
\providecommand \@ifxundefined [1]{%
 \@ifx{#1\undefined}
}%
\providecommand \@ifnum [1]{%
 \ifnum #1\expandafter \@firstoftwo
 \else \expandafter \@secondoftwo
 \fi
}%
\providecommand \@ifx [1]{%
 \ifx #1\expandafter \@firstoftwo
 \else \expandafter \@secondoftwo
 \fi
}%
\providecommand \natexlab [1]{#1}%
\providecommand \enquote  [1]{``#1''}%
\providecommand \bibnamefont  [1]{#1}%
\providecommand \bibfnamefont [1]{#1}%
\providecommand \citenamefont [1]{#1}%
\providecommand \href@noop [0]{\@secondoftwo}%
\providecommand \href [0]{\begingroup \@sanitize@url \@href}%
\providecommand \@href[1]{\@@startlink{#1}\@@href}%
\providecommand \@@href[1]{\endgroup#1\@@endlink}%
\providecommand \@sanitize@url [0]{\catcode `\\12\catcode `\$12\catcode `\&12\catcode `\#12\catcode `\^12\catcode `\_12\catcode `\%12\relax}%
\providecommand \@@startlink[1]{}%
\providecommand \@@endlink[0]{}%
\providecommand \url  [0]{\begingroup\@sanitize@url \@url }%
\providecommand \@url [1]{\endgroup\@href {#1}{\urlprefix }}%
\providecommand \urlprefix  [0]{URL }%
\providecommand \Eprint [0]{\href }%
\providecommand \doibase [0]{https://doi.org/}%
\providecommand \selectlanguage [0]{\@gobble}%
\providecommand \bibinfo  [0]{\@secondoftwo}%
\providecommand \bibfield  [0]{\@secondoftwo}%
\providecommand \translation [1]{[#1]}%
\providecommand \BibitemOpen [0]{}%
\providecommand \bibitemStop [0]{}%
\providecommand \bibitemNoStop [0]{.\EOS\space}%
\providecommand \EOS [0]{\spacefactor3000\relax}%
\providecommand \BibitemShut  [1]{\csname bibitem#1\endcsname}%
\let\auto@bib@innerbib\@empty
\bibitem [{\citenamefont {Kotera}\ and\ \citenamefont {Olinto}(2011)}]{Kotera:2011cp}%
  \BibitemOpen
  \bibfield  {author} {\bibinfo {author} {\bibfnamefont {K.}~\bibnamefont {Kotera}}\ and\ \bibinfo {author} {\bibfnamefont {A.~V.}\ \bibnamefont {Olinto}},\ }\bibfield  {title} {\bibinfo {title} {{The Astrophysics of Ultrahigh Energy Cosmic Rays}},\ }\href {https://doi.org/10.1146/annurev-astro-081710-102620} {\bibfield  {journal} {\bibinfo  {journal} {Ann. Rev. Astron. Astrophys.}\ }\textbf {\bibinfo {volume} {49}},\ \bibinfo {pages} {119} (\bibinfo {year} {2011})},\ \Eprint {https://arxiv.org/abs/1101.4256} {arXiv:1101.4256 [astro-ph.HE]} \BibitemShut {NoStop}%
\bibitem [{\citenamefont {Anchordoqui}(2019)}]{Anchordoqui:2018qom}%
  \BibitemOpen
  \bibfield  {author} {\bibinfo {author} {\bibfnamefont {L.~A.}\ \bibnamefont {Anchordoqui}},\ }\bibfield  {title} {\bibinfo {title} {{Ultra-High-Energy Cosmic Rays}},\ }\href {https://doi.org/10.1016/j.physrep.2019.01.002} {\bibfield  {journal} {\bibinfo  {journal} {Phys. Rept.}\ }\textbf {\bibinfo {volume} {801}},\ \bibinfo {pages} {1} (\bibinfo {year} {2019})},\ \Eprint {https://arxiv.org/abs/1807.09645} {arXiv:1807.09645 [astro-ph.HE]} \BibitemShut {NoStop}%
\bibitem [{\citenamefont {Alves~Batista}\ \emph {et~al.}(2019)\citenamefont {Alves~Batista} \emph {et~al.}}]{AlvesBatista:2019tlv}%
  \BibitemOpen
  \bibfield  {author} {\bibinfo {author} {\bibfnamefont {R.}~\bibnamefont {Alves~Batista}} \emph {et~al.},\ }\bibfield  {title} {\bibinfo {title} {{Open Questions in Cosmic-Ray Research at Ultrahigh Energies}},\ }\href {https://doi.org/10.3389/fspas.2019.00023} {\bibfield  {journal} {\bibinfo  {journal} {Front. Astron. Space Sci.}\ }\textbf {\bibinfo {volume} {6}},\ \bibinfo {pages} {23} (\bibinfo {year} {2019})},\ \Eprint {https://arxiv.org/abs/1903.06714} {arXiv:1903.06714 [astro-ph.HE]} \BibitemShut {NoStop}%
\bibitem [{\citenamefont {Waxman}(1995)}]{Waxman:1995vg}%
  \BibitemOpen
  \bibfield  {author} {\bibinfo {author} {\bibfnamefont {E.}~\bibnamefont {Waxman}},\ }\bibfield  {title} {\bibinfo {title} {{Cosmological gamma-ray bursts and the highest energy cosmic rays}},\ }\href {https://doi.org/10.1103/PhysRevLett.75.386} {\bibfield  {journal} {\bibinfo  {journal} {Phys. Rev. Lett.}\ }\textbf {\bibinfo {volume} {75}},\ \bibinfo {pages} {386} (\bibinfo {year} {1995})},\ \Eprint {https://arxiv.org/abs/astro-ph/9505082} {arXiv:astro-ph/9505082} \BibitemShut {NoStop}%
\bibitem [{\citenamefont {Vietri}(1995)}]{Vietri:1995hs}%
  \BibitemOpen
  \bibfield  {author} {\bibinfo {author} {\bibfnamefont {M.}~\bibnamefont {Vietri}},\ }\bibfield  {title} {\bibinfo {title} {{On the acceleration of ultrahigh-energy cosmic rays in gamma-ray bursts}},\ }\href {https://doi.org/10.1086/176448} {\bibfield  {journal} {\bibinfo  {journal} {Astrophys. J.}\ }\textbf {\bibinfo {volume} {453}},\ \bibinfo {pages} {883} (\bibinfo {year} {1995})},\ \Eprint {https://arxiv.org/abs/astro-ph/9506081} {arXiv:astro-ph/9506081} \BibitemShut {NoStop}%
\bibitem [{\citenamefont {Waxman}\ and\ \citenamefont {Bahcall}(1997)}]{Waxman:1997ti}%
  \BibitemOpen
  \bibfield  {author} {\bibinfo {author} {\bibfnamefont {E.}~\bibnamefont {Waxman}}\ and\ \bibinfo {author} {\bibfnamefont {J.~N.}\ \bibnamefont {Bahcall}},\ }\bibfield  {title} {\bibinfo {title} {{High-energy neutrinos from cosmological gamma-ray burst fireballs}},\ }\href {https://doi.org/10.1103/PhysRevLett.78.2292} {\bibfield  {journal} {\bibinfo  {journal} {Phys. Rev. Lett.}\ }\textbf {\bibinfo {volume} {78}},\ \bibinfo {pages} {2292} (\bibinfo {year} {1997})},\ \Eprint {https://arxiv.org/abs/astro-ph/9701231} {arXiv:astro-ph/9701231} \BibitemShut {NoStop}%
\bibitem [{\citenamefont {Woosley}(1993)}]{Woosley:1993wj}%
  \BibitemOpen
  \bibfield  {author} {\bibinfo {author} {\bibfnamefont {S.~E.}\ \bibnamefont {Woosley}},\ }\bibfield  {title} {\bibinfo {title} {{Gamma-ray bursts from stellar mass accretion disks around black holes}},\ }\href {https://doi.org/10.1086/172359} {\bibfield  {journal} {\bibinfo  {journal} {Astrophys. J.}\ }\textbf {\bibinfo {volume} {405}},\ \bibinfo {pages} {273} (\bibinfo {year} {1993})}\BibitemShut {NoStop}%
\bibitem [{\citenamefont {MacFadyen}\ and\ \citenamefont {Woosley}(1999)}]{MacFadyen:1998vz}%
  \BibitemOpen
  \bibfield  {author} {\bibinfo {author} {\bibfnamefont {A.}~\bibnamefont {MacFadyen}}\ and\ \bibinfo {author} {\bibfnamefont {S.~E.}\ \bibnamefont {Woosley}},\ }\bibfield  {title} {\bibinfo {title} {{Collapsars: Gamma-ray bursts and explosions in 'failed supernovae'}},\ }\href {https://doi.org/10.1086/307790} {\bibfield  {journal} {\bibinfo  {journal} {Astrophys. J.}\ }\textbf {\bibinfo {volume} {524}},\ \bibinfo {pages} {262} (\bibinfo {year} {1999})},\ \Eprint {https://arxiv.org/abs/astro-ph/9810274} {arXiv:astro-ph/9810274} \BibitemShut {NoStop}%
\bibitem [{\citenamefont {Piran}(2004)}]{Piran:2004ba}%
  \BibitemOpen
  \bibfield  {author} {\bibinfo {author} {\bibfnamefont {T.}~\bibnamefont {Piran}},\ }\bibfield  {title} {\bibinfo {title} {{The physics of gamma-ray bursts}},\ }\href {https://doi.org/10.1103/RevModPhys.76.1143} {\bibfield  {journal} {\bibinfo  {journal} {Rev. Mod. Phys.}\ }\textbf {\bibinfo {volume} {76}},\ \bibinfo {pages} {1143} (\bibinfo {year} {2004})},\ \Eprint {https://arxiv.org/abs/astro-ph/0405503} {arXiv:astro-ph/0405503} \BibitemShut {NoStop}%
\bibitem [{\citenamefont {Narayan}\ \emph {et~al.}(1992)\citenamefont {Narayan}, \citenamefont {Paczynski},\ and\ \citenamefont {Piran}}]{Narayan:1992iy}%
  \BibitemOpen
  \bibfield  {author} {\bibinfo {author} {\bibfnamefont {R.}~\bibnamefont {Narayan}}, \bibinfo {author} {\bibfnamefont {B.}~\bibnamefont {Paczynski}},\ and\ \bibinfo {author} {\bibfnamefont {T.}~\bibnamefont {Piran}},\ }\bibfield  {title} {\bibinfo {title} {{Gamma-ray bursts as the death throes of massive binary stars}},\ }\href {https://doi.org/10.1086/186493} {\bibfield  {journal} {\bibinfo  {journal} {Astrophys. J. Lett.}\ }\textbf {\bibinfo {volume} {395}},\ \bibinfo {pages} {L83} (\bibinfo {year} {1992})},\ \Eprint {https://arxiv.org/abs/astro-ph/9204001} {arXiv:astro-ph/9204001} \BibitemShut {NoStop}%
\bibitem [{\citenamefont {Fryer}\ \emph {et~al.}(1999)\citenamefont {Fryer}, \citenamefont {Benz}, \citenamefont {Herant},\ and\ \citenamefont {Colgate}}]{Fryer:1998jb}%
  \BibitemOpen
  \bibfield  {author} {\bibinfo {author} {\bibfnamefont {C.~L.}\ \bibnamefont {Fryer}}, \bibinfo {author} {\bibfnamefont {W.}~\bibnamefont {Benz}}, \bibinfo {author} {\bibfnamefont {M.}~\bibnamefont {Herant}},\ and\ \bibinfo {author} {\bibfnamefont {S.~A.}\ \bibnamefont {Colgate}},\ }\bibfield  {title} {\bibinfo {title} {{What can the accretion induced collapse of white dwarfs really explain?}},\ }\href {https://doi.org/10.1086/307119} {\bibfield  {journal} {\bibinfo  {journal} {Astrophys. J.}\ }\textbf {\bibinfo {volume} {516}},\ \bibinfo {pages} {892} (\bibinfo {year} {1999})},\ \Eprint {https://arxiv.org/abs/astro-ph/9812058} {arXiv:astro-ph/9812058} \BibitemShut {NoStop}%
\bibitem [{\citenamefont {Metzger}\ \emph {et~al.}(2008{\natexlab{a}})\citenamefont {Metzger}, \citenamefont {Quataert},\ and\ \citenamefont {Thompson}}]{Metzger:2007cd}%
  \BibitemOpen
  \bibfield  {author} {\bibinfo {author} {\bibfnamefont {B.~D.}\ \bibnamefont {Metzger}}, \bibinfo {author} {\bibfnamefont {E.}~\bibnamefont {Quataert}},\ and\ \bibinfo {author} {\bibfnamefont {T.~A.}\ \bibnamefont {Thompson}},\ }\bibfield  {title} {\bibinfo {title} {{Short Duration Gamma-Ray Bursts with Extended Emission from Proto-Magnetar Spin-Down}},\ }\href {https://doi.org/10.1111/j.1365-2966.2008.12923.x} {\bibfield  {journal} {\bibinfo  {journal} {Mon. Not. Roy. Astron. Soc.}\ }\textbf {\bibinfo {volume} {385}},\ \bibinfo {pages} {1455} (\bibinfo {year} {2008}{\natexlab{a}})},\ \Eprint {https://arxiv.org/abs/0712.1233} {arXiv:0712.1233 [astro-ph]} \BibitemShut {NoStop}%
\bibitem [{\citenamefont {Chen}\ \emph {et~al.}(2024)\citenamefont {Chen}, \citenamefont {Shen}, \citenamefont {Tan}, \citenamefont {Wang}, \citenamefont {Xiong}, \citenamefont {Chen},\ and\ \citenamefont {Zhang}}]{Chen:2024arq}%
  \BibitemOpen
  \bibfield  {author} {\bibinfo {author} {\bibfnamefont {J.}~\bibnamefont {Chen}}, \bibinfo {author} {\bibfnamefont {R.-F.}\ \bibnamefont {Shen}}, \bibinfo {author} {\bibfnamefont {W.-J.}\ \bibnamefont {Tan}}, \bibinfo {author} {\bibfnamefont {C.-W.}\ \bibnamefont {Wang}}, \bibinfo {author} {\bibfnamefont {S.-L.}\ \bibnamefont {Xiong}}, \bibinfo {author} {\bibfnamefont {R.-C.}\ \bibnamefont {Chen}},\ and\ \bibinfo {author} {\bibfnamefont {B.-B.}\ \bibnamefont {Zhang}},\ }\bibfield  {title} {\bibinfo {title} {{Repeated Partial Disruptions in a White Dwarf{\textendash}Neutron Star or White Dwarf{\textendash}Black Hole Merger Modulate the Prompt Emission of Long-duration Merger-type GRBs}},\ }\href {https://doi.org/10.3847/2041-8213/ad7737} {\bibfield  {journal} {\bibinfo  {journal} {Astrophys. J. Lett.}\ }\textbf {\bibinfo {volume} {973}},\ \bibinfo {pages} {L33} (\bibinfo {year} {2024})},\ \Eprint {https://arxiv.org/abs/2409.00472} {arXiv:2409.00472 [astro-ph.HE]} \BibitemShut {NoStop}%
\bibitem [{\citenamefont {Cheong}\ \emph {et~al.}(2025)\citenamefont {Cheong}, \citenamefont {Pitik}, \citenamefont {Longo~Micchi},\ and\ \citenamefont {Radice}}]{Cheong:2024hrd}%
  \BibitemOpen
  \bibfield  {author} {\bibinfo {author} {\bibfnamefont {P.~C.-K.}\ \bibnamefont {Cheong}}, \bibinfo {author} {\bibfnamefont {T.}~\bibnamefont {Pitik}}, \bibinfo {author} {\bibfnamefont {L.~F.}\ \bibnamefont {Longo~Micchi}},\ and\ \bibinfo {author} {\bibfnamefont {D.}~\bibnamefont {Radice}},\ }\bibfield  {title} {\bibinfo {title} {{Gamma-Ray Bursts and Kilonovae from the Accretion-induced Collapse of White Dwarfs}},\ }\href {https://doi.org/10.3847/2041-8213/ada1cc} {\bibfield  {journal} {\bibinfo  {journal} {Astrophys. J. Lett.}\ }\textbf {\bibinfo {volume} {978}},\ \bibinfo {pages} {L38} (\bibinfo {year} {2025})},\ \Eprint {https://arxiv.org/abs/2410.10938} {arXiv:2410.10938 [astro-ph.HE]} \BibitemShut {NoStop}%
\bibitem [{\citenamefont {Combi}\ \emph {et~al.}(2025)\citenamefont {Combi}, \citenamefont {Siegel},\ and\ \citenamefont {Metzger}}]{Combi:2025yvs}%
  \BibitemOpen
  \bibfield  {author} {\bibinfo {author} {\bibfnamefont {L.}~\bibnamefont {Combi}}, \bibinfo {author} {\bibfnamefont {D.~M.}\ \bibnamefont {Siegel}},\ and\ \bibinfo {author} {\bibfnamefont {B.~D.}\ \bibnamefont {Metzger}},\ }\bibfield  {title} {\bibinfo {title} {{Jet-driven explosion of an accretion-induced white-dwarf collapse via a magnetorotational dynamo}},\ }\href@noop {} {\  (\bibinfo {year} {2025})},\ \Eprint {https://arxiv.org/abs/2509.19799} {arXiv:2509.19799 [astro-ph.HE]} \BibitemShut {NoStop}%
\bibitem [{\citenamefont {Nomoto}(1986)}]{NOMOTO1986249}%
  \BibitemOpen
  \bibfield  {author} {\bibinfo {author} {\bibfnamefont {K.}~\bibnamefont {Nomoto}},\ }\bibfield  {title} {\bibinfo {title} {The fate of accreting white dwarfs: Type i supernovae vs. collapse},\ }\href {https://doi.org/https://doi.org/10.1016/0146-6410(86)90020-7} {\bibfield  {journal} {\bibinfo  {journal} {Progress in Particle and Nuclear Physics}\ }\textbf {\bibinfo {volume} {17}},\ \bibinfo {pages} {249} (\bibinfo {year} {1986})}\BibitemShut {NoStop}%
\bibitem [{\citenamefont {Yoon}\ \emph {et~al.}(2007)\citenamefont {Yoon}, \citenamefont {Podsiadlowski},\ and\ \citenamefont {Rosswog}}]{Yoon:2007pw}%
  \BibitemOpen
  \bibfield  {author} {\bibinfo {author} {\bibfnamefont {S.-C.}\ \bibnamefont {Yoon}}, \bibinfo {author} {\bibfnamefont {P.}~\bibnamefont {Podsiadlowski}},\ and\ \bibinfo {author} {\bibfnamefont {S.}~\bibnamefont {Rosswog}},\ }\bibfield  {title} {\bibinfo {title} {{Remnant evolution after a carbon-oxygen white dwarf merger}},\ }\href {https://doi.org/10.1111/j.1365-2966.2007.12161.x} {\bibfield  {journal} {\bibinfo  {journal} {Mon. Not. Roy. Astron. Soc.}\ }\textbf {\bibinfo {volume} {380}},\ \bibinfo {pages} {933} (\bibinfo {year} {2007})},\ \Eprint {https://arxiv.org/abs/0704.0297} {arXiv:0704.0297 [astro-ph]} \BibitemShut {NoStop}%
\bibitem [{\citenamefont {Chandrasekhar}(1931)}]{Chandrasekhar:1931ih}%
  \BibitemOpen
  \bibfield  {author} {\bibinfo {author} {\bibfnamefont {S.}~\bibnamefont {Chandrasekhar}},\ }\bibfield  {title} {\bibinfo {title} {{The maximum mass of ideal white dwarfs}},\ }\href {https://doi.org/10.1086/143324} {\bibfield  {journal} {\bibinfo  {journal} {Astrophys. J.}\ }\textbf {\bibinfo {volume} {74}},\ \bibinfo {pages} {81} (\bibinfo {year} {1931})}\BibitemShut {NoStop}%
\bibitem [{\citenamefont {{Nomoto}}\ and\ \citenamefont {{Kondo}}(1991)}]{1991ApJ...367L..19N}%
  \BibitemOpen
  \bibfield  {author} {\bibinfo {author} {\bibfnamefont {K.}~\bibnamefont {{Nomoto}}}\ and\ \bibinfo {author} {\bibfnamefont {Y.}~\bibnamefont {{Kondo}}},\ }\bibfield  {title} {\bibinfo {title} {{Conditions for Accretion-induced Collapse of White Dwarfs}},\ }\href {https://doi.org/10.1086/185922} {\bibfield  {journal} {\bibinfo  {journal} {Astrophys. J. Lett.}\ }\textbf {\bibinfo {volume} {367}},\ \bibinfo {pages} {L19} (\bibinfo {year} {1991})}\BibitemShut {NoStop}%
\bibitem [{\citenamefont {Ablimit}\ and\ \citenamefont {Li}(2015)}]{Ablimit:2014vka}%
  \BibitemOpen
  \bibfield  {author} {\bibinfo {author} {\bibfnamefont {I.}~\bibnamefont {Ablimit}}\ and\ \bibinfo {author} {\bibfnamefont {X.-D.}\ \bibnamefont {Li}},\ }\bibfield  {title} {\bibinfo {title} {{Formation of Binary Millisecond Pulsars by Accretion-Induced Collapse of White Dwarfs under Wind-Driven Evolution}},\ }\href {https://doi.org/10.1088/0004-637X/800/2/98} {\bibfield  {journal} {\bibinfo  {journal} {Astrophys. J.}\ }\textbf {\bibinfo {volume} {800}},\ \bibinfo {pages} {98} (\bibinfo {year} {2015})},\ \Eprint {https://arxiv.org/abs/1412.7245} {arXiv:1412.7245 [astro-ph.HE]} \BibitemShut {NoStop}%
\bibitem [{\citenamefont {Wang}\ \emph {et~al.}(2022)\citenamefont {Wang}, \citenamefont {Liu},\ and\ \citenamefont {Chen}}]{Wang:2022duk}%
  \BibitemOpen
  \bibfield  {author} {\bibinfo {author} {\bibfnamefont {B.}~\bibnamefont {Wang}}, \bibinfo {author} {\bibfnamefont {D.}~\bibnamefont {Liu}},\ and\ \bibinfo {author} {\bibfnamefont {H.}~\bibnamefont {Chen}},\ }\bibfield  {title} {\bibinfo {title} {{Formation of millisecond pulsars with long orbital periods by accretion-induced collapse of white dwarfs}},\ }\href {https://doi.org/10.1093/mnras/stac114} {\bibfield  {journal} {\bibinfo  {journal} {Mon. Not. Roy. Astron. Soc.}\ }\textbf {\bibinfo {volume} {510}},\ \bibinfo {pages} {6011} (\bibinfo {year} {2022})},\ \Eprint {https://arxiv.org/abs/2201.03827} {arXiv:2201.03827 [astro-ph.SR]} \BibitemShut {NoStop}%
\bibitem [{\citenamefont {Piersanti}\ \emph {et~al.}(2003)\citenamefont {Piersanti}, \citenamefont {Gagliardi}, \citenamefont {Iben},\ and\ \citenamefont {Tornambe}}]{Piersanti:2002sb}%
  \BibitemOpen
  \bibfield  {author} {\bibinfo {author} {\bibfnamefont {L.}~\bibnamefont {Piersanti}}, \bibinfo {author} {\bibfnamefont {S.}~\bibnamefont {Gagliardi}}, \bibinfo {author} {\bibfnamefont {I.}~\bibnamefont {Iben}, \bibfnamefont {Jr.}},\ and\ \bibinfo {author} {\bibfnamefont {A.}~\bibnamefont {Tornambe}},\ }\bibfield  {title} {\bibinfo {title} {{Carbon - oxygen white dwarfs accreting CO-rich matter I: A Comparison between rotating and non-rotating models}},\ }\href {https://doi.org/10.1086/345444} {\bibfield  {journal} {\bibinfo  {journal} {Astrophys. J.}\ }\textbf {\bibinfo {volume} {583}},\ \bibinfo {pages} {885} (\bibinfo {year} {2003})},\ \Eprint {https://arxiv.org/abs/astro-ph/0210624} {arXiv:astro-ph/0210624} \BibitemShut {NoStop}%
\bibitem [{\citenamefont {{Piersanti}}\ \emph {et~al.}(2003)\citenamefont {{Piersanti}}, \citenamefont {{Gagliardi}}, \citenamefont {{Iben}},\ and\ \citenamefont {{Tornamb{\'e}}}}]{2003ApJ...598.1229P}%
  \BibitemOpen
  \bibfield  {author} {\bibinfo {author} {\bibfnamefont {L.}~\bibnamefont {{Piersanti}}}, \bibinfo {author} {\bibfnamefont {S.}~\bibnamefont {{Gagliardi}}}, \bibinfo {author} {\bibfnamefont {I.}~\bibnamefont {{Iben}}, \bibfnamefont {Jr.}},\ and\ \bibinfo {author} {\bibfnamefont {A.}~\bibnamefont {{Tornamb{\'e}}}},\ }\bibfield  {title} {\bibinfo {title} {{Carbon-Oxygen White Dwarf Accreting CO-Rich Matter. II. Self-Regulating Accretion Process up to the Explosive Stage}},\ }\href {https://doi.org/10.1086/378952} {\bibfield  {journal} {\bibinfo  {journal} {"Astrophys. J. Lett."}\ }\textbf {\bibinfo {volume} {598}},\ \bibinfo {pages} {1229} (\bibinfo {year} {2003})}\BibitemShut {NoStop}%
\bibitem [{\citenamefont {Uenishi}\ \emph {et~al.}(2003)\citenamefont {Uenishi}, \citenamefont {Nomoto},\ and\ \citenamefont {Hachisu}}]{Uenishi:2003sk}%
  \BibitemOpen
  \bibfield  {author} {\bibinfo {author} {\bibfnamefont {T.}~\bibnamefont {Uenishi}}, \bibinfo {author} {\bibfnamefont {K.}~\bibnamefont {Nomoto}},\ and\ \bibinfo {author} {\bibfnamefont {I.}~\bibnamefont {Hachisu}},\ }\bibfield  {title} {\bibinfo {title} {{Evolution of rotating accreting white dwarfs and the diversity of type Ia supernovae}},\ }\href {https://doi.org/10.1086/377248} {\bibfield  {journal} {\bibinfo  {journal} {Astrophys. J.}\ }\textbf {\bibinfo {volume} {595}},\ \bibinfo {pages} {1094} (\bibinfo {year} {2003})},\ \Eprint {https://arxiv.org/abs/astro-ph/0309433} {arXiv:astro-ph/0309433} \BibitemShut {NoStop}%
\bibitem [{\citenamefont {Saio}\ and\ \citenamefont {Nomoto}(2004)}]{Saio:2004gz}%
  \BibitemOpen
  \bibfield  {author} {\bibinfo {author} {\bibfnamefont {H.}~\bibnamefont {Saio}}\ and\ \bibinfo {author} {\bibfnamefont {K.}~\bibnamefont {Nomoto}},\ }\bibfield  {title} {\bibinfo {title} {{Off - center carbon ignition in rapidly rotating, accreting carbon - oxygen white dwarfs}},\ }\href {https://doi.org/10.1086/423976} {\bibfield  {journal} {\bibinfo  {journal} {Astrophys. J.}\ }\textbf {\bibinfo {volume} {615}},\ \bibinfo {pages} {444} (\bibinfo {year} {2004})},\ \Eprint {https://arxiv.org/abs/astro-ph/0401141} {arXiv:astro-ph/0401141} \BibitemShut {NoStop}%
\bibitem [{\citenamefont {{Kashyap}}\ \emph {et~al.}(2018)\citenamefont {{Kashyap}}, \citenamefont {{Haque}}, \citenamefont {{Lor{\'e}n-Aguilar}}, \citenamefont {{Garc{\'\i}a-Berro}},\ and\ \citenamefont {{Fisher}}}]{2018ApJ...869..140K}%
  \BibitemOpen
  \bibfield  {author} {\bibinfo {author} {\bibfnamefont {R.}~\bibnamefont {{Kashyap}}}, \bibinfo {author} {\bibfnamefont {T.}~\bibnamefont {{Haque}}}, \bibinfo {author} {\bibfnamefont {P.}~\bibnamefont {{Lor{\'e}n-Aguilar}}}, \bibinfo {author} {\bibfnamefont {E.}~\bibnamefont {{Garc{\'\i}a-Berro}}},\ and\ \bibinfo {author} {\bibfnamefont {R.}~\bibnamefont {{Fisher}}},\ }\bibfield  {title} {\bibinfo {title} {{Double-degenerate Carbon-Oxygen and Oxygen-Neon White Dwarf Mergers: A New Mechanism for Faint and Rapid Type Ia Supernovae}},\ }\href {https://doi.org/10.3847/1538-4357/aaedb7} {\bibfield  {journal} {\bibinfo  {journal} {"Astrophys. J."}\ }\textbf {\bibinfo {volume} {869}},\ \bibinfo {eid} {140} (\bibinfo {year} {2018})},\ \Eprint {https://arxiv.org/abs/1811.00013} {arXiv:1811.00013 [astro-ph.SR]} \BibitemShut {NoStop}%
\bibitem [{\citenamefont {{Ferrario}}\ \emph {et~al.}(2015)\citenamefont {{Ferrario}}, \citenamefont {{de Martino}},\ and\ \citenamefont {{G{\"a}nsicke}}}]{2015SSRv..191..111F}%
  \BibitemOpen
  \bibfield  {author} {\bibinfo {author} {\bibfnamefont {L.}~\bibnamefont {{Ferrario}}}, \bibinfo {author} {\bibfnamefont {D.}~\bibnamefont {{de Martino}}},\ and\ \bibinfo {author} {\bibfnamefont {B.~T.}\ \bibnamefont {{G{\"a}nsicke}}},\ }\bibfield  {title} {\bibinfo {title} {{Magnetic White Dwarfs}},\ }\href {https://doi.org/10.1007/s11214-015-0152-0} {\bibfield  {journal} {\bibinfo  {journal} {"Space Science Reviews"}\ }\textbf {\bibinfo {volume} {191}},\ \bibinfo {pages} {111} (\bibinfo {year} {2015})},\ \Eprint {https://arxiv.org/abs/1504.08072} {arXiv:1504.08072 [astro-ph.SR]} \BibitemShut {NoStop}%
\bibitem [{\citenamefont {{Zhu}}\ \emph {et~al.}(2015)\citenamefont {{Zhu}}, \citenamefont {{Pakmor}}, \citenamefont {{van Kerkwijk}},\ and\ \citenamefont {{Chang}}}]{2015ApJ...806L...1Z}%
  \BibitemOpen
  \bibfield  {author} {\bibinfo {author} {\bibfnamefont {C.}~\bibnamefont {{Zhu}}}, \bibinfo {author} {\bibfnamefont {R.}~\bibnamefont {{Pakmor}}}, \bibinfo {author} {\bibfnamefont {M.~H.}\ \bibnamefont {{van Kerkwijk}}},\ and\ \bibinfo {author} {\bibfnamefont {P.}~\bibnamefont {{Chang}}},\ }\bibfield  {title} {\bibinfo {title} {{Magnetized Moving Mesh Merger of a Carbon-Oxygen White Dwarf Binary}},\ }\href {https://doi.org/10.1088/2041-8205/806/1/L1} {\bibfield  {journal} {\bibinfo  {journal} {"Astrophys. J. Lett."}\ }\textbf {\bibinfo {volume} {806}},\ \bibinfo {eid} {L1} (\bibinfo {year} {2015})},\ \Eprint {https://arxiv.org/abs/1504.01732} {arXiv:1504.01732 [astro-ph.SR]} \BibitemShut {NoStop}%
\bibitem [{\citenamefont {{Ferrario}}\ \emph {et~al.}(2020)\citenamefont {{Ferrario}}, \citenamefont {{Wickramasinghe}},\ and\ \citenamefont {{Kawka}}}]{2020AdSpR..66.1025F}%
  \BibitemOpen
  \bibfield  {author} {\bibinfo {author} {\bibfnamefont {L.}~\bibnamefont {{Ferrario}}}, \bibinfo {author} {\bibfnamefont {D.}~\bibnamefont {{Wickramasinghe}}},\ and\ \bibinfo {author} {\bibfnamefont {A.}~\bibnamefont {{Kawka}}},\ }\bibfield  {title} {\bibinfo {title} {{Magnetic fields in isolated and interacting white dwarfs}},\ }\href {https://doi.org/10.1016/j.asr.2019.11.012} {\bibfield  {journal} {\bibinfo  {journal} {Advances in Space Research}\ }\textbf {\bibinfo {volume} {66}},\ \bibinfo {pages} {1025} (\bibinfo {year} {2020})},\ \Eprint {https://arxiv.org/abs/2001.10147} {arXiv:2001.10147 [astro-ph.SR]} \BibitemShut {NoStop}%
\bibitem [{\citenamefont {Pakmor}\ \emph {et~al.}(2024)\citenamefont {Pakmor} \emph {et~al.}}]{Pakmor:2024efc}%
  \BibitemOpen
  \bibfield  {author} {\bibinfo {author} {\bibfnamefont {R.}~\bibnamefont {Pakmor}} \emph {et~al.},\ }\bibfield  {title} {\bibinfo {title} {{Large-scale ordered magnetic fields generated in mergers of helium white dwarfs}},\ }\href {https://doi.org/10.1051/0004-6361/202451352} {\bibfield  {journal} {\bibinfo  {journal} {Astron. Astrophys.}\ }\textbf {\bibinfo {volume} {691}},\ \bibinfo {pages} {A179} (\bibinfo {year} {2024})},\ \Eprint {https://arxiv.org/abs/2407.02566} {arXiv:2407.02566 [astro-ph.SR]} \BibitemShut {NoStop}%
\bibitem [{\citenamefont {Yi}\ and\ \citenamefont {Blackman}(1998)}]{Yi:1997qb}%
  \BibitemOpen
  \bibfield  {author} {\bibinfo {author} {\bibfnamefont {I.}~\bibnamefont {Yi}}\ and\ \bibinfo {author} {\bibfnamefont {E.~G.}\ \bibnamefont {Blackman}},\ }\bibfield  {title} {\bibinfo {title} {{An Explanation for the bimodal duration distribution of gamma-ray bursts: millisecond puslars from accretion induced collapse}},\ }\href {https://doi.org/10.1086/311192} {\bibfield  {journal} {\bibinfo  {journal} {Astrophys. J. Lett.}\ }\textbf {\bibinfo {volume} {494}},\ \bibinfo {pages} {L163} (\bibinfo {year} {1998})},\ \Eprint {https://arxiv.org/abs/astro-ph/9710149} {arXiv:astro-ph/9710149} \BibitemShut {NoStop}%
\bibitem [{\citenamefont {Perley}\ \emph {et~al.}(2009)\citenamefont {Perley} \emph {et~al.}}]{Perley:2008ay}%
  \BibitemOpen
  \bibfield  {author} {\bibinfo {author} {\bibfnamefont {D.~A.}\ \bibnamefont {Perley}} \emph {et~al.},\ }\bibfield  {title} {\bibinfo {title} {{GRB 080503: Implications of a Naked Short Gamma-Ray Burst Dominated by Extended Emission}},\ }\href {https://doi.org/10.1088/0004-637X/696/2/1871} {\bibfield  {journal} {\bibinfo  {journal} {Astrophys. J.}\ }\textbf {\bibinfo {volume} {696}},\ \bibinfo {pages} {1871} (\bibinfo {year} {2009})},\ \Eprint {https://arxiv.org/abs/0811.1044} {arXiv:0811.1044 [astro-ph]} \BibitemShut {NoStop}%
\bibitem [{\citenamefont {Longo~Micchi}\ \emph {et~al.}(2023)\citenamefont {Longo~Micchi}, \citenamefont {Radice},\ and\ \citenamefont {Chirenti}}]{LongoMicchi:2023khv}%
  \BibitemOpen
  \bibfield  {author} {\bibinfo {author} {\bibfnamefont {L.~F.}\ \bibnamefont {Longo~Micchi}}, \bibinfo {author} {\bibfnamefont {D.}~\bibnamefont {Radice}},\ and\ \bibinfo {author} {\bibfnamefont {C.}~\bibnamefont {Chirenti}},\ }\bibfield  {title} {\bibinfo {title} {{Multimessenger emission from the accretion-induced collapse of white dwarfs}},\ }\href {https://doi.org/10.1093/mnras/stad2420} {\bibfield  {journal} {\bibinfo  {journal} {Mon. Not. Roy. Astron. Soc.}\ }\textbf {\bibinfo {volume} {525}},\ \bibinfo {pages} {6359} (\bibinfo {year} {2023})},\ \Eprint {https://arxiv.org/abs/2306.04711} {arXiv:2306.04711 [astro-ph.HE]} \BibitemShut {NoStop}%
\bibitem [{\citenamefont {Batziou}\ \emph {et~al.}(2025)\citenamefont {Batziou}, \citenamefont {Glas}, \citenamefont {Janka}, \citenamefont {Ehring}, \citenamefont {Abdikamalov},\ and\ \citenamefont {Just}}]{Batziou:2024ory}%
  \BibitemOpen
  \bibfield  {author} {\bibinfo {author} {\bibfnamefont {E.}~\bibnamefont {Batziou}}, \bibinfo {author} {\bibfnamefont {R.}~\bibnamefont {Glas}}, \bibinfo {author} {\bibfnamefont {H.~T.}\ \bibnamefont {Janka}}, \bibinfo {author} {\bibfnamefont {J.}~\bibnamefont {Ehring}}, \bibinfo {author} {\bibfnamefont {E.}~\bibnamefont {Abdikamalov}},\ and\ \bibinfo {author} {\bibfnamefont {O.}~\bibnamefont {Just}},\ }\bibfield  {title} {\bibinfo {title} {{Nucleosynthesis Conditions in Outflows of White Dwarfs Collapsing to Neutron Stars}},\ }\href {https://doi.org/10.3847/1538-4357/adc300} {\bibfield  {journal} {\bibinfo  {journal} {Astrophys. J.}\ }\textbf {\bibinfo {volume} {984}},\ \bibinfo {pages} {197} (\bibinfo {year} {2025})},\ \Eprint {https://arxiv.org/abs/2412.02756} {arXiv:2412.02756 [astro-ph.HE]} \BibitemShut {NoStop}%
\bibitem [{\citenamefont {Yip}\ \emph {et~al.}(2024)\citenamefont {Yip}, \citenamefont {Chu}, \citenamefont {Leung},\ and\ \citenamefont {Lin}}]{Yip:2024akb}%
  \BibitemOpen
  \bibfield  {author} {\bibinfo {author} {\bibfnamefont {C.-M.}\ \bibnamefont {Yip}}, \bibinfo {author} {\bibfnamefont {M.-C.}\ \bibnamefont {Chu}}, \bibinfo {author} {\bibfnamefont {S.-C.}\ \bibnamefont {Leung}},\ and\ \bibinfo {author} {\bibfnamefont {L.-M.}\ \bibnamefont {Lin}},\ }\bibfield  {title} {\bibinfo {title} {{On the Nucleosynthesis in Accretion-Induced Collapse of White Dwarfs}},\ }\href@noop {} {\  (\bibinfo {year} {2024})},\ \Eprint {https://arxiv.org/abs/2401.03798} {arXiv:2401.03798 [astro-ph.HE]} \BibitemShut {NoStop}%
\bibitem [{\citenamefont {Pitik}\ \emph {et~al.}(2026{\natexlab{a}})\citenamefont {Pitik}, \citenamefont {Radice}, \citenamefont {Kasen}, \citenamefont {Magistrelli}, \citenamefont {Cheong},\ and\ \citenamefont {Bernuzzi}}]{Pitik:2026bjm}%
  \BibitemOpen
  \bibfield  {author} {\bibinfo {author} {\bibfnamefont {T.}~\bibnamefont {Pitik}}, \bibinfo {author} {\bibfnamefont {D.}~\bibnamefont {Radice}}, \bibinfo {author} {\bibfnamefont {D.}~\bibnamefont {Kasen}}, \bibinfo {author} {\bibfnamefont {F.}~\bibnamefont {Magistrelli}}, \bibinfo {author} {\bibfnamefont {P.~C.-K.}\ \bibnamefont {Cheong}},\ and\ \bibinfo {author} {\bibfnamefont {S.}~\bibnamefont {Bernuzzi}},\ }\bibfield  {title} {\bibinfo {title} {{Collapse of Magnetized White Dwarfs as site of Heavy Element Formation and Kilonova Signal}},\ }\href@noop {} {\  (\bibinfo {year} {2026}{\natexlab{a}})},\ \Eprint {https://arxiv.org/abs/2602.21291} {arXiv:2602.21291 [astro-ph.HE]} \BibitemShut {NoStop}%
\bibitem [{\citenamefont {Pitik}\ \emph {et~al.}(2026{\natexlab{b}})\citenamefont {Pitik}, \citenamefont {Qian}, \citenamefont {Radice},\ and\ \citenamefont {Kasen}}]{Pitik:2026qfj}%
  \BibitemOpen
  \bibfield  {author} {\bibinfo {author} {\bibfnamefont {T.}~\bibnamefont {Pitik}}, \bibinfo {author} {\bibfnamefont {Y.-Z.}\ \bibnamefont {Qian}}, \bibinfo {author} {\bibfnamefont {D.}~\bibnamefont {Radice}},\ and\ \bibinfo {author} {\bibfnamefont {D.}~\bibnamefont {Kasen}},\ }\bibfield  {title} {\bibinfo {title} {{Gamma-ray Signatures of r-Process Radioactivity from the Collapse of Magnetized White Dwarfs}},\ }\href@noop {} {\  (\bibinfo {year} {2026}{\natexlab{b}})},\ \Eprint {https://arxiv.org/abs/2603.08792} {arXiv:2603.08792 [astro-ph.HE]} \BibitemShut {NoStop}%
\bibitem [{\citenamefont {de~Gouveia Dal~Pino}\ and\ \citenamefont {Lazarian}(2000)}]{deGouveiaDalPino:2000jz}%
  \BibitemOpen
  \bibfield  {author} {\bibinfo {author} {\bibfnamefont {E.}~\bibnamefont {de~Gouveia Dal~Pino}}\ and\ \bibinfo {author} {\bibfnamefont {A.}~\bibnamefont {Lazarian}},\ }\bibfield  {title} {\bibinfo {title} {{Ultrahigh-energy cosmic ray acceleration by magnetic reconnection in newborn accretion induced collapse pulsars}},\ }\href {https://doi.org/10.1086/312730} {\bibfield  {journal} {\bibinfo  {journal} {Astrophys. J. Lett.}\ }\textbf {\bibinfo {volume} {536}},\ \bibinfo {pages} {L31} (\bibinfo {year} {2000})},\ \Eprint {https://arxiv.org/abs/astro-ph/0002155} {arXiv:astro-ph/0002155} \BibitemShut {NoStop}%
\bibitem [{\citenamefont {De~Gouveia Dal~Pino}\ and\ \citenamefont {Lazarian}(2001)}]{DeGouveiaDalPino:2001ci}%
  \BibitemOpen
  \bibfield  {author} {\bibinfo {author} {\bibfnamefont {E.~M.}\ \bibnamefont {De~Gouveia Dal~Pino}}\ and\ \bibinfo {author} {\bibfnamefont {A.}~\bibnamefont {Lazarian}},\ }\bibfield  {title} {\bibinfo {title} {{Constraints on the acceleration of ultrahigh-energy cosmic rays in accretion induced collapse pulsars}},\ }\href {https://doi.org/10.1086/322509} {\bibfield  {journal} {\bibinfo  {journal} {Astrophys. J.}\ }\textbf {\bibinfo {volume} {560}},\ \bibinfo {pages} {358} (\bibinfo {year} {2001})},\ \Eprint {https://arxiv.org/abs/astro-ph/0106452} {arXiv:astro-ph/0106452} \BibitemShut {NoStop}%
\bibitem [{\citenamefont {{Dar}}\ \emph {et~al.}(1992)\citenamefont {{Dar}}, \citenamefont {{Kozlovsky}}, \citenamefont {{Nussinov}},\ and\ \citenamefont {{Ramaty}}}]{1992ApJ...388..164D}%
  \BibitemOpen
  \bibfield  {author} {\bibinfo {author} {\bibfnamefont {A.}~\bibnamefont {{Dar}}}, \bibinfo {author} {\bibfnamefont {B.~Z.}\ \bibnamefont {{Kozlovsky}}}, \bibinfo {author} {\bibfnamefont {S.}~\bibnamefont {{Nussinov}}},\ and\ \bibinfo {author} {\bibfnamefont {R.}~\bibnamefont {{Ramaty}}},\ }\bibfield  {title} {\bibinfo {title} {{Gamma-Ray Bursts and Cosmic Rays from Accretion-induced Collapse}},\ }\href {https://doi.org/10.1086/171138} {\bibfield  {journal} {\bibinfo  {journal} {\apj}\ }\textbf {\bibinfo {volume} {388}},\ \bibinfo {pages} {164} (\bibinfo {year} {1992})}\BibitemShut {NoStop}%
\bibitem [{\citenamefont {Mei}\ \emph {et~al.}(2022)\citenamefont {Mei} \emph {et~al.}}]{Mei:2022ncd}%
  \BibitemOpen
  \bibfield  {author} {\bibinfo {author} {\bibfnamefont {A.}~\bibnamefont {Mei}} \emph {et~al.},\ }\bibfield  {title} {\bibinfo {title} {{Gigaelectronvolt emission from a compact binary merger}},\ }\href {https://doi.org/10.1038/s41586-022-05404-7} {\bibfield  {journal} {\bibinfo  {journal} {Nature}\ }\textbf {\bibinfo {volume} {612}},\ \bibinfo {pages} {236} (\bibinfo {year} {2022})},\ \Eprint {https://arxiv.org/abs/2205.08566} {arXiv:2205.08566 [astro-ph.HE]} \BibitemShut {NoStop}%
\bibitem [{\citenamefont {Rastinejad}\ \emph {et~al.}(2022)\citenamefont {Rastinejad} \emph {et~al.}}]{Rastinejad:2022zbg}%
  \BibitemOpen
  \bibfield  {author} {\bibinfo {author} {\bibfnamefont {J.~C.}\ \bibnamefont {Rastinejad}} \emph {et~al.},\ }\bibfield  {title} {\bibinfo {title} {{A kilonova following a long-duration gamma-ray burst at 350 Mpc}},\ }\href {https://doi.org/10.1038/s41586-022-05390-w} {\bibfield  {journal} {\bibinfo  {journal} {Nature}\ }\textbf {\bibinfo {volume} {612}},\ \bibinfo {pages} {223} (\bibinfo {year} {2022})},\ \Eprint {https://arxiv.org/abs/2204.10864} {arXiv:2204.10864 [astro-ph.HE]} \BibitemShut {NoStop}%
\bibitem [{\citenamefont {Troja}\ \emph {et~al.}(2022)\citenamefont {Troja} \emph {et~al.}}]{Troja:2022yya}%
  \BibitemOpen
  \bibfield  {author} {\bibinfo {author} {\bibfnamefont {E.}~\bibnamefont {Troja}} \emph {et~al.},\ }\bibfield  {title} {\bibinfo {title} {{A nearby long gamma-ray burst from a merger of compact objects}},\ }\href {https://doi.org/10.1038/s41586-022-05327-3} {\bibfield  {journal} {\bibinfo  {journal} {Nature}\ }\textbf {\bibinfo {volume} {612}},\ \bibinfo {pages} {228} (\bibinfo {year} {2022})},\ \Eprint {https://arxiv.org/abs/2209.03363} {arXiv:2209.03363 [astro-ph.HE]} \BibitemShut {NoStop}%
\bibitem [{\citenamefont {Yang}\ \emph {et~al.}(2022)\citenamefont {Yang}, \citenamefont {Ai}, \citenamefont {Zhang}, \citenamefont {Zhang}, \citenamefont {Liu}, \citenamefont {Wang}, \citenamefont {Yang}, \citenamefont {Yin}, \citenamefont {Li},\ and\ \citenamefont {L{\"u}}}]{Yang:2022qmy}%
  \BibitemOpen
  \bibfield  {author} {\bibinfo {author} {\bibfnamefont {J.}~\bibnamefont {Yang}}, \bibinfo {author} {\bibfnamefont {S.}~\bibnamefont {Ai}}, \bibinfo {author} {\bibfnamefont {B.-B.}\ \bibnamefont {Zhang}}, \bibinfo {author} {\bibfnamefont {B.}~\bibnamefont {Zhang}}, \bibinfo {author} {\bibfnamefont {Z.-K.}\ \bibnamefont {Liu}}, \bibinfo {author} {\bibfnamefont {X.~I.}\ \bibnamefont {Wang}}, \bibinfo {author} {\bibfnamefont {Y.-H.}\ \bibnamefont {Yang}}, \bibinfo {author} {\bibfnamefont {Y.-H.}\ \bibnamefont {Yin}}, \bibinfo {author} {\bibfnamefont {Y.}~\bibnamefont {Li}},\ and\ \bibinfo {author} {\bibfnamefont {H.-J.}\ \bibnamefont {L{\"u}}},\ }\bibfield  {title} {\bibinfo {title} {{A long-duration gamma-ray burst with a peculiar origin}},\ }\href {https://doi.org/10.1038/s41586-022-05403-8} {\bibfield  {journal} {\bibinfo  {journal} {Nature}\ }\textbf {\bibinfo {volume} {612}},\ \bibinfo {pages} {232} (\bibinfo {year} {2022})},\ \Eprint {https://arxiv.org/abs/2204.12771} {arXiv:2204.12771 [astro-ph.HE]} \BibitemShut {NoStop}%
\bibitem [{\citenamefont {{Burrows}}\ \emph {et~al.}(2023)\citenamefont {{Burrows}}, \citenamefont {{Gropp}}, \citenamefont {{Osborne}}, \citenamefont {{Page}}, \citenamefont {{D'Elia}}, \citenamefont {{Sbarufatti}}, \citenamefont {{D'Ai}}, \citenamefont {{Dichiara}}, \citenamefont {{Evans}},\ and\ \citenamefont {{Swift-XRT Team}}}]{2023GCN.33429....1B}%
  \BibitemOpen
  \bibfield  {author} {\bibinfo {author} {\bibfnamefont {D.~N.}\ \bibnamefont {{Burrows}}}, \bibinfo {author} {\bibfnamefont {J.~D.}\ \bibnamefont {{Gropp}}}, \bibinfo {author} {\bibfnamefont {J.~P.}\ \bibnamefont {{Osborne}}}, \bibinfo {author} {\bibfnamefont {K.~L.}\ \bibnamefont {{Page}}}, \bibinfo {author} {\bibfnamefont {V.}~\bibnamefont {{D'Elia}}}, \bibinfo {author} {\bibfnamefont {B.}~\bibnamefont {{Sbarufatti}}}, \bibinfo {author} {\bibfnamefont {A.}~\bibnamefont {{D'Ai}}}, \bibinfo {author} {\bibfnamefont {S.}~\bibnamefont {{Dichiara}}}, \bibinfo {author} {\bibfnamefont {P.~A.}\ \bibnamefont {{Evans}}},\ and\ \bibinfo {author} {\bibnamefont {{Swift-XRT Team}}},\ }\bibfield  {title} {\bibinfo {title} {{GRB 230307A: Swift-XRT observations}},\ }\href@noop {} {\bibfield  {journal} {\bibinfo  {journal} {GRB Coordinates Network}\ }\textbf {\bibinfo {volume} {33429}},\ \bibinfo {pages} {1} (\bibinfo {year} {2023})}\BibitemShut {NoStop}%
\bibitem [{\citenamefont {Dalessi}\ \emph {et~al.}(2025)\citenamefont {Dalessi} \emph {et~al.}}]{Dalessi:2025unu}%
  \BibitemOpen
  \bibfield  {author} {\bibinfo {author} {\bibfnamefont {S.}~\bibnamefont {Dalessi}} \emph {et~al.},\ }\bibfield  {title} {\bibinfo {title} {{Fermi-GBM Observations of GRB 230307A: An Exceptionally Bright Long-duration Gamma-ray Burst with an Associated Kilonova}},\ }\href {https://doi.org/10.3847/1538-4357/ae0a1d} {\bibfield  {journal} {\bibinfo  {journal} {Astrophys. J.}\ }\textbf {\bibinfo {volume} {994}},\ \bibinfo {pages} {17} (\bibinfo {year} {2025})},\ \Eprint {https://arxiv.org/abs/2507.12637} {arXiv:2507.12637 [astro-ph.HE]} \BibitemShut {NoStop}%
\bibitem [{\citenamefont {Levan}\ \emph {et~al.}(2024)\citenamefont {Levan} \emph {et~al.}}]{JWST:2023jqa}%
  \BibitemOpen
  \bibfield  {author} {\bibinfo {author} {\bibfnamefont {A.~J.}\ \bibnamefont {Levan}} \emph {et~al.} (\bibinfo {collaboration} {JWST}),\ }\bibfield  {title} {\bibinfo {title} {{Heavy-element production in a compact object merger observed by JWST}},\ }\href {https://doi.org/10.1038/s41586-023-06759-1} {\bibfield  {journal} {\bibinfo  {journal} {Nature}\ }\textbf {\bibinfo {volume} {626}},\ \bibinfo {pages} {737} (\bibinfo {year} {2024})},\ \Eprint {https://arxiv.org/abs/2307.02098} {arXiv:2307.02098 [astro-ph.HE]} \BibitemShut {NoStop}%
\bibitem [{\citenamefont {Tomsick}\ \emph {et~al.}(2023)\citenamefont {Tomsick} \emph {et~al.}}]{Tomsick:2023aue}%
  \BibitemOpen
  \bibfield  {author} {\bibinfo {author} {\bibfnamefont {J.~A.}\ \bibnamefont {Tomsick}} \emph {et~al.},\ }\bibfield  {title} {\bibinfo {title} {{The Compton Spectrometer and Imager}},\ }\href {https://doi.org/10.22323/1.444.0745} {\bibfield  {journal} {\bibinfo  {journal} {PoS}\ }\textbf {\bibinfo {volume} {ICRC2023}},\ \bibinfo {pages} {745} (\bibinfo {year} {2023})},\ \Eprint {https://arxiv.org/abs/2308.12362} {arXiv:2308.12362 [astro-ph.HE]} \BibitemShut {NoStop}%
\bibitem [{\citenamefont {Caputo}\ \emph {et~al.}(2022)\citenamefont {Caputo} \emph {et~al.}}]{Caputo:2022xpx}%
  \BibitemOpen
  \bibfield  {author} {\bibinfo {author} {\bibfnamefont {R.}~\bibnamefont {Caputo}} \emph {et~al.},\ }\bibfield  {title} {\bibinfo {title} {{All-sky Medium Energy Gamma-ray Observatory eXplorer mission concept}},\ }\href {https://doi.org/10.1117/1.JATIS.8.4.044003} {\bibfield  {journal} {\bibinfo  {journal} {J. Astron. Telesc. Instrum. Syst.}\ }\textbf {\bibinfo {volume} {8}},\ \bibinfo {pages} {044003} (\bibinfo {year} {2022})},\ \Eprint {https://arxiv.org/abs/2208.04990} {arXiv:2208.04990 [astro-ph.IM]} \BibitemShut {NoStop}%
\bibitem [{\citenamefont {Metzger}\ \emph {et~al.}(2011{\natexlab{a}})\citenamefont {Metzger}, \citenamefont {Giannios}, \citenamefont {Thompson}, \citenamefont {Bucciantini},\ and\ \citenamefont {Quataert}}]{Metzger:2010pp}%
  \BibitemOpen
  \bibfield  {author} {\bibinfo {author} {\bibfnamefont {B.~D.}\ \bibnamefont {Metzger}}, \bibinfo {author} {\bibfnamefont {D.}~\bibnamefont {Giannios}}, \bibinfo {author} {\bibfnamefont {T.~A.}\ \bibnamefont {Thompson}}, \bibinfo {author} {\bibfnamefont {N.}~\bibnamefont {Bucciantini}},\ and\ \bibinfo {author} {\bibfnamefont {E.}~\bibnamefont {Quataert}},\ }\bibfield  {title} {\bibinfo {title} {{The Proto-Magnetar Model for Gamma-Ray Bursts}},\ }\href {https://doi.org/10.1111/j.1365-2966.2011.18280.x} {\bibfield  {journal} {\bibinfo  {journal} {Mon. Not. Roy. Astron. Soc.}\ }\textbf {\bibinfo {volume} {413}},\ \bibinfo {pages} {2031} (\bibinfo {year} {2011}{\natexlab{a}})},\ \Eprint {https://arxiv.org/abs/1012.0001} {arXiv:1012.0001 [astro-ph.HE]} \BibitemShut {NoStop}%
\bibitem [{\citenamefont {Metzger}\ \emph {et~al.}(2011{\natexlab{b}})\citenamefont {Metzger}, \citenamefont {Giannios},\ and\ \citenamefont {Horiuchi}}]{Metzger:2011xs}%
  \BibitemOpen
  \bibfield  {author} {\bibinfo {author} {\bibfnamefont {B.~D.}\ \bibnamefont {Metzger}}, \bibinfo {author} {\bibfnamefont {D.}~\bibnamefont {Giannios}},\ and\ \bibinfo {author} {\bibfnamefont {S.}~\bibnamefont {Horiuchi}},\ }\bibfield  {title} {\bibinfo {title} {{Heavy Nuclei Synthesized in Gamma-Ray Burst Outflows as the Source of UHECRs}},\ }\href {https://doi.org/10.1111/j.1365-2966.2011.18873.x} {\bibfield  {journal} {\bibinfo  {journal} {Mon. Not. Roy. Astron. Soc.}\ }\textbf {\bibinfo {volume} {415}},\ \bibinfo {pages} {2495} (\bibinfo {year} {2011}{\natexlab{b}})},\ \Eprint {https://arxiv.org/abs/1101.4019} {arXiv:1101.4019 [astro-ph.HE]} \BibitemShut {NoStop}%
\bibitem [{\citenamefont {Qian}\ and\ \citenamefont {Woosley}(1996)}]{Qian:1996xt}%
  \BibitemOpen
  \bibfield  {author} {\bibinfo {author} {\bibfnamefont {Y.~Z.}\ \bibnamefont {Qian}}\ and\ \bibinfo {author} {\bibfnamefont {S.~E.}\ \bibnamefont {Woosley}},\ }\bibfield  {title} {\bibinfo {title} {{Nucleosynthesis in neutrino driven winds: 1. The Physical conditions}},\ }\href {https://doi.org/10.1086/177973} {\bibfield  {journal} {\bibinfo  {journal} {Astrophys. J.}\ }\textbf {\bibinfo {volume} {471}},\ \bibinfo {pages} {331} (\bibinfo {year} {1996})},\ \Eprint {https://arxiv.org/abs/astro-ph/9611094} {arXiv:astro-ph/9611094} \BibitemShut {NoStop}%
\bibitem [{\citenamefont {Zhang}\ and\ \citenamefont {Meszaros}(2004)}]{Zhang:2003uk}%
  \BibitemOpen
  \bibfield  {author} {\bibinfo {author} {\bibfnamefont {B.}~\bibnamefont {Zhang}}\ and\ \bibinfo {author} {\bibfnamefont {P.}~\bibnamefont {Meszaros}},\ }\bibfield  {title} {\bibinfo {title} {{Gamma-ray bursts: Progress, problems {\&} prospects}},\ }\href {https://doi.org/10.1142/S0217751X0401746X} {\bibfield  {journal} {\bibinfo  {journal} {Int. J. Mod. Phys. A}\ }\textbf {\bibinfo {volume} {19}},\ \bibinfo {pages} {2385} (\bibinfo {year} {2004})},\ \Eprint {https://arxiv.org/abs/astro-ph/0311321} {arXiv:astro-ph/0311321} \BibitemShut {NoStop}%
\bibitem [{\citenamefont {Meszaros}(2006)}]{Meszaros:2006rc}%
  \BibitemOpen
  \bibfield  {author} {\bibinfo {author} {\bibfnamefont {P.}~\bibnamefont {Meszaros}},\ }\bibfield  {title} {\bibinfo {title} {{Gamma-Ray Bursts}},\ }\href {https://doi.org/10.1088/0034-4885/69/8/R01} {\bibfield  {journal} {\bibinfo  {journal} {Rept. Prog. Phys.}\ }\textbf {\bibinfo {volume} {69}},\ \bibinfo {pages} {2259} (\bibinfo {year} {2006})},\ \Eprint {https://arxiv.org/abs/astro-ph/0605208} {arXiv:astro-ph/0605208} \BibitemShut {NoStop}%
\bibitem [{\citenamefont {Lyubarsky}(2005)}]{Lyubarsky:2005zt}%
  \BibitemOpen
  \bibfield  {author} {\bibinfo {author} {\bibfnamefont {Y.~E.}\ \bibnamefont {Lyubarsky}},\ }\bibfield  {title} {\bibinfo {title} {{On the relativistic magnetic reconnection}},\ }\href {https://doi.org/10.1111/j.1365-2966.2005.08767.x} {\bibfield  {journal} {\bibinfo  {journal} {Mon. Not. Roy. Astron. Soc.}\ }\textbf {\bibinfo {volume} {358}},\ \bibinfo {pages} {113} (\bibinfo {year} {2005})},\ \Eprint {https://arxiv.org/abs/astro-ph/0501392} {arXiv:astro-ph/0501392} \BibitemShut {NoStop}%
\bibitem [{\citenamefont {{Drake}}\ \emph {et~al.}(2009)\citenamefont {{Drake}}, \citenamefont {{Cassak}}, \citenamefont {{Shay}}, \citenamefont {{Swisdak}},\ and\ \citenamefont {{Quataert}}}]{2009ApJ...700L..16D}%
  \BibitemOpen
  \bibfield  {author} {\bibinfo {author} {\bibfnamefont {J.~F.}\ \bibnamefont {{Drake}}}, \bibinfo {author} {\bibfnamefont {P.~A.}\ \bibnamefont {{Cassak}}}, \bibinfo {author} {\bibfnamefont {M.~A.}\ \bibnamefont {{Shay}}}, \bibinfo {author} {\bibfnamefont {M.}~\bibnamefont {{Swisdak}}},\ and\ \bibinfo {author} {\bibfnamefont {E.}~\bibnamefont {{Quataert}}},\ }\bibfield  {title} {\bibinfo {title} {{A Magnetic Reconnection Mechanism for Ion Acceleration and Abundance Enhancements in Impulsive Flares}},\ }\href {https://doi.org/10.1088/0004-637X/700/1/L16} {\bibfield  {journal} {\bibinfo  {journal} {Astrophys. J. Lett.}\ }\textbf {\bibinfo {volume} {700}},\ \bibinfo {pages} {L16} (\bibinfo {year} {2009})}\BibitemShut {NoStop}%
\bibitem [{\citenamefont {Sironi}\ and\ \citenamefont {Spitkovsky}(2014)}]{Sironi:2014jfa}%
  \BibitemOpen
  \bibfield  {author} {\bibinfo {author} {\bibfnamefont {L.}~\bibnamefont {Sironi}}\ and\ \bibinfo {author} {\bibfnamefont {A.}~\bibnamefont {Spitkovsky}},\ }\bibfield  {title} {\bibinfo {title} {{Relativistic Reconnection: an Efficient Source of Non-Thermal Particles}},\ }\href {https://doi.org/10.1088/2041-8205/783/1/L21} {\bibfield  {journal} {\bibinfo  {journal} {Astrophys. J. Lett.}\ }\textbf {\bibinfo {volume} {783}},\ \bibinfo {pages} {L21} (\bibinfo {year} {2014})},\ \Eprint {https://arxiv.org/abs/1401.5471} {arXiv:1401.5471 [astro-ph.HE]} \BibitemShut {NoStop}%
\bibitem [{\citenamefont {Murase}\ \emph {et~al.}(2008)\citenamefont {Murase}, \citenamefont {Ioka}, \citenamefont {Nagataki},\ and\ \citenamefont {Nakamura}}]{Murase:2008mr}%
  \BibitemOpen
  \bibfield  {author} {\bibinfo {author} {\bibfnamefont {K.}~\bibnamefont {Murase}}, \bibinfo {author} {\bibfnamefont {K.}~\bibnamefont {Ioka}}, \bibinfo {author} {\bibfnamefont {S.}~\bibnamefont {Nagataki}},\ and\ \bibinfo {author} {\bibfnamefont {T.}~\bibnamefont {Nakamura}},\ }\bibfield  {title} {\bibinfo {title} {{High-energy cosmic-ray nuclei from high- and low-luminosity gamma-ray bursts and implications for multi-messenger astronomy}},\ }\href {https://doi.org/10.1103/PhysRevD.78.023005} {\bibfield  {journal} {\bibinfo  {journal} {Phys. Rev. D}\ }\textbf {\bibinfo {volume} {78}},\ \bibinfo {pages} {023005} (\bibinfo {year} {2008})},\ \Eprint {https://arxiv.org/abs/0801.2861} {arXiv:0801.2861 [astro-ph]} \BibitemShut {NoStop}%
\bibitem [{\citenamefont {Fixsen}(2009)}]{Fixsen:2009ug}%
  \BibitemOpen
  \bibfield  {author} {\bibinfo {author} {\bibfnamefont {D.~J.}\ \bibnamefont {Fixsen}},\ }\bibfield  {title} {\bibinfo {title} {{The Temperature of the Cosmic Microwave Background}},\ }\href {https://doi.org/10.1088/0004-637X/707/2/916} {\bibfield  {journal} {\bibinfo  {journal} {Astrophys. J.}\ }\textbf {\bibinfo {volume} {707}},\ \bibinfo {pages} {916} (\bibinfo {year} {2009})},\ \Eprint {https://arxiv.org/abs/0911.1955} {arXiv:0911.1955 [astro-ph.CO]} \BibitemShut {NoStop}%
\bibitem [{\citenamefont {Allard}\ \emph {et~al.}(2005)\citenamefont {Allard}, \citenamefont {Parizot}, \citenamefont {Khan}, \citenamefont {Goriely},\ and\ \citenamefont {Olinto}}]{Allard:2005ha}%
  \BibitemOpen
  \bibfield  {author} {\bibinfo {author} {\bibfnamefont {D.}~\bibnamefont {Allard}}, \bibinfo {author} {\bibfnamefont {E.}~\bibnamefont {Parizot}}, \bibinfo {author} {\bibfnamefont {E.}~\bibnamefont {Khan}}, \bibinfo {author} {\bibfnamefont {S.}~\bibnamefont {Goriely}},\ and\ \bibinfo {author} {\bibfnamefont {A.~V.}\ \bibnamefont {Olinto}},\ }\bibfield  {title} {\bibinfo {title} {{UHE nuclei propagation and the interpretation of the ankle in the cosmic-ray spectrum}},\ }\href {https://doi.org/10.1051/0004-6361:200500199} {\bibfield  {journal} {\bibinfo  {journal} {Astron. Astrophys.}\ }\textbf {\bibinfo {volume} {443}},\ \bibinfo {pages} {L29} (\bibinfo {year} {2005})},\ \Eprint {https://arxiv.org/abs/astro-ph/0505566} {arXiv:astro-ph/0505566} \BibitemShut {NoStop}%
\bibitem [{\citenamefont {{Conselice}}\ \emph {et~al.}(2016)\citenamefont {{Conselice}}, \citenamefont {{Wilkinson}}, \citenamefont {{Duncan}},\ and\ \citenamefont {{Mortlock}}}]{2016ApJ...830...83C}%
  \BibitemOpen
  \bibfield  {author} {\bibinfo {author} {\bibfnamefont {C.~J.}\ \bibnamefont {{Conselice}}}, \bibinfo {author} {\bibfnamefont {A.}~\bibnamefont {{Wilkinson}}}, \bibinfo {author} {\bibfnamefont {K.}~\bibnamefont {{Duncan}}},\ and\ \bibinfo {author} {\bibfnamefont {A.}~\bibnamefont {{Mortlock}}},\ }\bibfield  {title} {\bibinfo {title} {{The Evolution of Galaxy Number Density at z < 8 and Its Implications}},\ }\href {https://doi.org/10.3847/0004-637X/830/2/83} {\bibfield  {journal} {\bibinfo  {journal} {\apj}\ }\textbf {\bibinfo {volume} {830}},\ \bibinfo {eid} {83} (\bibinfo {year} {2016})},\ \Eprint {https://arxiv.org/abs/1607.03909} {arXiv:1607.03909 [astro-ph.GA]} \BibitemShut {NoStop}%
\bibitem [{\citenamefont {Dessart}\ \emph {et~al.}(2006)\citenamefont {Dessart}, \citenamefont {Burrows}, \citenamefont {Ott}, \citenamefont {Livne}, \citenamefont {Yoon},\ and\ \citenamefont {Langer}}]{Dessart:2006gd}%
  \BibitemOpen
  \bibfield  {author} {\bibinfo {author} {\bibfnamefont {L.}~\bibnamefont {Dessart}}, \bibinfo {author} {\bibfnamefont {A.}~\bibnamefont {Burrows}}, \bibinfo {author} {\bibfnamefont {C.}~\bibnamefont {Ott}}, \bibinfo {author} {\bibfnamefont {E.}~\bibnamefont {Livne}}, \bibinfo {author} {\bibfnamefont {S.-C.}\ \bibnamefont {Yoon}},\ and\ \bibinfo {author} {\bibfnamefont {N.}~\bibnamefont {Langer}},\ }\bibfield  {title} {\bibinfo {title} {{Multi-dimensional simulations of the accretion-induced collapse of white dwarfs to neutron stars}},\ }\href {https://doi.org/10.1086/503626} {\bibfield  {journal} {\bibinfo  {journal} {Astrophys. J.}\ }\textbf {\bibinfo {volume} {644}},\ \bibinfo {pages} {1063} (\bibinfo {year} {2006})},\ \Eprint {https://arxiv.org/abs/astro-ph/0601603} {arXiv:astro-ph/0601603} \BibitemShut {NoStop}%
\bibitem [{\citenamefont {Ruiter}\ \emph {et~al.}(2019)\citenamefont {Ruiter}, \citenamefont {Ferrario}, \citenamefont {Belczynski}, \citenamefont {Seitenzahl}, \citenamefont {Crocker},\ and\ \citenamefont {Karakas}}]{Ruiter:2018ouw}%
  \BibitemOpen
  \bibfield  {author} {\bibinfo {author} {\bibfnamefont {A.~J.}\ \bibnamefont {Ruiter}}, \bibinfo {author} {\bibfnamefont {L.}~\bibnamefont {Ferrario}}, \bibinfo {author} {\bibfnamefont {K.}~\bibnamefont {Belczynski}}, \bibinfo {author} {\bibfnamefont {I.~R.}\ \bibnamefont {Seitenzahl}}, \bibinfo {author} {\bibfnamefont {R.~M.}\ \bibnamefont {Crocker}},\ and\ \bibinfo {author} {\bibfnamefont {A.~I.}\ \bibnamefont {Karakas}},\ }\bibfield  {title} {\bibinfo {title} {{On the formation of neutron stars via accretion-induced collapse in binaries}},\ }\href {https://doi.org/10.1093/mnras/stz001} {\bibfield  {journal} {\bibinfo  {journal} {Mon. Not. Roy. Astron. Soc.}\ }\textbf {\bibinfo {volume} {484}},\ \bibinfo {pages} {698} (\bibinfo {year} {2019})},\ \Eprint {https://arxiv.org/abs/1802.02437} {arXiv:1802.02437 [astro-ph.SR]} \BibitemShut {NoStop}%
\bibitem [{\citenamefont {Murase}\ and\ \citenamefont {Takami}(2009)}]{Murase:2008sa}%
  \BibitemOpen
  \bibfield  {author} {\bibinfo {author} {\bibfnamefont {K.}~\bibnamefont {Murase}}\ and\ \bibinfo {author} {\bibfnamefont {H.}~\bibnamefont {Takami}},\ }\bibfield  {title} {\bibinfo {title} {{Implications of Ultra-High-Energy Cosmic Rays for Transient Sources in the Auger Era}},\ }\href {https://doi.org/10.1088/0004-637X/690/1/L14} {\bibfield  {journal} {\bibinfo  {journal} {Astrophys. J. Lett.}\ }\textbf {\bibinfo {volume} {690}},\ \bibinfo {pages} {L14} (\bibinfo {year} {2009})},\ \Eprint {https://arxiv.org/abs/0810.1813} {arXiv:0810.1813 [astro-ph]} \BibitemShut {NoStop}%
\bibitem [{\citenamefont {Desai}\ \emph {et~al.}(2026)\citenamefont {Desai}, \citenamefont {Combi}, \citenamefont {Siegel},\ and\ \citenamefont {Metzger}}]{Desai:2026llu}%
  \BibitemOpen
  \bibfield  {author} {\bibinfo {author} {\bibfnamefont {D.~K.}\ \bibnamefont {Desai}}, \bibinfo {author} {\bibfnamefont {L.}~\bibnamefont {Combi}}, \bibinfo {author} {\bibfnamefont {D.~M.}\ \bibnamefont {Siegel}},\ and\ \bibinfo {author} {\bibfnamefont {B.~D.}\ \bibnamefont {Metzger}},\ }\bibfield  {title} {\bibinfo {title} {{Relativistic jets from millisecond proto-magnetars}},\ }\href@noop {} {\  (\bibinfo {year} {2026})},\ \Eprint {https://arxiv.org/abs/2601.07918} {arXiv:2601.07918 [astro-ph.HE]} \BibitemShut {NoStop}%
\bibitem [{\citenamefont {Nakazato}\ \emph {et~al.}(2018)\citenamefont {Nakazato}, \citenamefont {Suzuki},\ and\ \citenamefont {Togashi}}]{Nakazato:2017ynw}%
  \BibitemOpen
  \bibfield  {author} {\bibinfo {author} {\bibfnamefont {K.}~\bibnamefont {Nakazato}}, \bibinfo {author} {\bibfnamefont {H.}~\bibnamefont {Suzuki}},\ and\ \bibinfo {author} {\bibfnamefont {H.}~\bibnamefont {Togashi}},\ }\bibfield  {title} {\bibinfo {title} {{Heavy nuclei as thermal insulation for protoneutron stars}},\ }\href {https://doi.org/10.1103/PhysRevC.97.035804} {\bibfield  {journal} {\bibinfo  {journal} {Phys. Rev. C}\ }\textbf {\bibinfo {volume} {97}},\ \bibinfo {pages} {035804} (\bibinfo {year} {2018})},\ \Eprint {https://arxiv.org/abs/1710.10441} {arXiv:1710.10441 [astro-ph.HE]} \BibitemShut {NoStop}%
\bibitem [{\citenamefont {Suwa}\ \emph {et~al.}(2019)\citenamefont {Suwa}, \citenamefont {Sumiyoshi}, \citenamefont {Nakazato}, \citenamefont {Takahira}, \citenamefont {Koshio}, \citenamefont {Mori},\ and\ \citenamefont {Wendell}}]{Suwa:2019svl}%
  \BibitemOpen
  \bibfield  {author} {\bibinfo {author} {\bibfnamefont {Y.}~\bibnamefont {Suwa}}, \bibinfo {author} {\bibfnamefont {K.}~\bibnamefont {Sumiyoshi}}, \bibinfo {author} {\bibfnamefont {K.}~\bibnamefont {Nakazato}}, \bibinfo {author} {\bibfnamefont {Y.}~\bibnamefont {Takahira}}, \bibinfo {author} {\bibfnamefont {Y.}~\bibnamefont {Koshio}}, \bibinfo {author} {\bibfnamefont {M.}~\bibnamefont {Mori}},\ and\ \bibinfo {author} {\bibfnamefont {R.~A.}\ \bibnamefont {Wendell}},\ }\bibfield  {title} {\bibinfo {title} {{Observing Supernova Neutrino Light Curves with Super-Kamiokande: Expected Event Number over 10 s}},\ }\href {https://doi.org/10.3847/1538-4357/ab2e05} {\bibfield  {journal} {\bibinfo  {journal} {Astrophys. J.}\ }\textbf {\bibinfo {volume} {881}},\ \bibinfo {pages} {139} (\bibinfo {year} {2019})},\ \Eprint {https://arxiv.org/abs/1904.09996} {arXiv:1904.09996 [astro-ph.HE]} \BibitemShut {NoStop}%
\bibitem [{\citenamefont {Fiorillo}\ \emph {et~al.}(2023)\citenamefont {Fiorillo}, \citenamefont {Heinlein}, \citenamefont {Janka}, \citenamefont {Raffelt}, \citenamefont {Vitagliano},\ and\ \citenamefont {Bollig}}]{Fiorillo:2023frv}%
  \BibitemOpen
  \bibfield  {author} {\bibinfo {author} {\bibfnamefont {D.~F.~G.}\ \bibnamefont {Fiorillo}}, \bibinfo {author} {\bibfnamefont {M.}~\bibnamefont {Heinlein}}, \bibinfo {author} {\bibfnamefont {H.-T.}\ \bibnamefont {Janka}}, \bibinfo {author} {\bibfnamefont {G.}~\bibnamefont {Raffelt}}, \bibinfo {author} {\bibfnamefont {E.}~\bibnamefont {Vitagliano}},\ and\ \bibinfo {author} {\bibfnamefont {R.}~\bibnamefont {Bollig}},\ }\bibfield  {title} {\bibinfo {title} {{Supernova simulations confront SN 1987A neutrinos}},\ }\href {https://doi.org/10.1103/PhysRevD.108.083040} {\bibfield  {journal} {\bibinfo  {journal} {Phys. Rev. D}\ }\textbf {\bibinfo {volume} {108}},\ \bibinfo {pages} {083040} (\bibinfo {year} {2023})},\ \Eprint {https://arxiv.org/abs/2308.01403} {arXiv:2308.01403 [astro-ph.HE]} \BibitemShut {NoStop}%
\bibitem [{\citenamefont {Lucente}\ \emph {et~al.}(2024)\citenamefont {Lucente}, \citenamefont {Heinlein}, \citenamefont {Janka},\ and\ \citenamefont {Mirizzi}}]{Lucente:2024ngp}%
  \BibitemOpen
  \bibfield  {author} {\bibinfo {author} {\bibfnamefont {G.}~\bibnamefont {Lucente}}, \bibinfo {author} {\bibfnamefont {M.}~\bibnamefont {Heinlein}}, \bibinfo {author} {\bibfnamefont {H.-T.}\ \bibnamefont {Janka}},\ and\ \bibinfo {author} {\bibfnamefont {A.}~\bibnamefont {Mirizzi}},\ }\bibfield  {title} {\bibinfo {title} {{Simple fits for the neutrino luminosities from protoneutron star cooling}},\ }\href {https://doi.org/10.1103/PhysRevD.110.063023} {\bibfield  {journal} {\bibinfo  {journal} {Phys. Rev. D}\ }\textbf {\bibinfo {volume} {110}},\ \bibinfo {pages} {063023} (\bibinfo {year} {2024})},\ \Eprint {https://arxiv.org/abs/2405.00769} {arXiv:2405.00769 [astro-ph.HE]} \BibitemShut {NoStop}%
\bibitem [{\citenamefont {Thompson}\ \emph {et~al.}(2004)\citenamefont {Thompson}, \citenamefont {Chang},\ and\ \citenamefont {Quataert}}]{Thompson:2004wi}%
  \BibitemOpen
  \bibfield  {author} {\bibinfo {author} {\bibfnamefont {T.~A.}\ \bibnamefont {Thompson}}, \bibinfo {author} {\bibfnamefont {P.}~\bibnamefont {Chang}},\ and\ \bibinfo {author} {\bibfnamefont {E.}~\bibnamefont {Quataert}},\ }\bibfield  {title} {\bibinfo {title} {{Magnetar spindown, hyper-energetic supernovae, and gamma ray bursts}},\ }\href {https://doi.org/10.1086/421969} {\bibfield  {journal} {\bibinfo  {journal} {Astrophys. J.}\ }\textbf {\bibinfo {volume} {611}},\ \bibinfo {pages} {380} (\bibinfo {year} {2004})},\ \Eprint {https://arxiv.org/abs/astro-ph/0401555} {arXiv:astro-ph/0401555} \BibitemShut {NoStop}%
\bibitem [{\citenamefont {Metzger}\ \emph {et~al.}(2008{\natexlab{b}})\citenamefont {Metzger}, \citenamefont {Thompson},\ and\ \citenamefont {Quataert}}]{Metzger:2007kj}%
  \BibitemOpen
  \bibfield  {author} {\bibinfo {author} {\bibfnamefont {B.~D.}\ \bibnamefont {Metzger}}, \bibinfo {author} {\bibfnamefont {T.~A.}\ \bibnamefont {Thompson}},\ and\ \bibinfo {author} {\bibfnamefont {E.}~\bibnamefont {Quataert}},\ }\bibfield  {title} {\bibinfo {title} {{On the Conditions for Neutron-Rich Gamma-Ray Burst Outflows}},\ }\href {https://doi.org/10.1086/526418} {\bibfield  {journal} {\bibinfo  {journal} {Astrophys. J.}\ }\textbf {\bibinfo {volume} {676}},\ \bibinfo {pages} {1130} (\bibinfo {year} {2008}{\natexlab{b}})},\ \Eprint {https://arxiv.org/abs/0708.3395} {arXiv:0708.3395 [astro-ph]} \BibitemShut {NoStop}%
\bibitem [{\citenamefont {Goldreich}\ and\ \citenamefont {Julian}(1969)}]{Goldreich:1969sb}%
  \BibitemOpen
  \bibfield  {author} {\bibinfo {author} {\bibfnamefont {P.}~\bibnamefont {Goldreich}}\ and\ \bibinfo {author} {\bibfnamefont {W.~H.}\ \bibnamefont {Julian}},\ }\bibfield  {title} {\bibinfo {title} {{Pulsar electrodynamics}},\ }\href {https://doi.org/10.1086/150119} {\bibfield  {journal} {\bibinfo  {journal} {Astrophys. J.}\ }\textbf {\bibinfo {volume} {157}},\ \bibinfo {pages} {869} (\bibinfo {year} {1969})}\BibitemShut {NoStop}%
\end{thebibliography}%
\clearpage
\newpage
\appendix
\onecolumngrid
\maketitle
\onecolumngrid
\begin{center}
	\textbf{\Large Supplementary Material}
	 \bigskip\\
		\textbf{\large Ultra high-energy cosmic rays from relativistic outflows in accretion induced collapse of white dwarfs}
		 \medskip\\
   {Mainak Mukhopadhyay and Shunsaku Horiuchi}
\end{center}
\renewcommand{\thesection}{S\arabic{section}}
\renewcommand{\theequation}{S\arabic{equation}}
\renewcommand{\thefigure}{S\arabic{figure}}
\renewcommand{\thetable}{S\arabic{table}}
\renewcommand{\thepage}{S\arabic{page}}
\setcounter{equation}{0}
\setcounter{figure}{0}
\setcounter{table}{0}
\setcounter{page}{1}
\setcounter{secnumdepth}{4}
\section{On the mass loss rate}
\label{appsec:massloss_rate}
After the collapse of a WD, the resulting PNS loses mass as a result of neutrino heating. The rate of mass-loss for the PNS is crucial for evaluating the resulting magnetization of the wind (see Eq.~\ref{eq:sigma0}). In this section, we provide details on the mass-loss rate of a typical pulsar born as a result of an AIC.

The Kelvin-Helmholtz cooling phase of the PNS lasts roughly $\sim 10$ s. In this phase, the PNS loses its gravitational binding energy in the form of neutrino emission in all-flavors. The resulting neutrino flux leads to weak interactions that transfer energy from the neutrinos to the matter around the surface of the neutron star and results in heating. This process, known as neutrino heating, causes the material to expand away from the PNS and form a mass outflow, which is defined as the \emph{neutrino-driven wind}.

The dominant weak interactions include the absorption of $\nu_e$ ($\nu_e + n \leftrightarrow e^- + p$), $\bar{\nu}_e$ ($\bar{\nu}_e + p \leftrightarrow e^+ + n$), neutrino-electron scattering, and neutrino antineutrino annihilations producing pairs. Since this phase is well past the accretion phase when convective instabilities dominate the surroundings, a quasi-steady state approximation in the gain region can be reasonably assumed for our purposes, although more non-trivial dynamics are indeed possible. So, the mass-loss rate we evaluate is under the assumption of a spherically symmetric outflow. As evident from the above equations, the outflow comprises photons, relativistic $e^+-e^-$ pairs, and non-relativistic free protons and neutrons. But, it is dominated mostly by pairs and photons\footnote{Note that the mass loading relevant for $\dot{M}$ is however baryonic.}. This is because the nucleons have subsonic velocities, need at least the gravitational potential energy ($\sim G M_* m_n/R_* \sim 200$ MeV, where $m_n$ is the mass of the nucleon and $\sim 1$ GeV) to move away from the vicinity of the PNS as an outflow, and the surface temperature of the PNS and the thermal kinetic energy of the nucleons are $\sim$ MeV. Therefore, most of the energy from the neutrino flux goes to the pairs and the photons, which is indeed the case for the above mentioned weak interactions.

Our estimates assume hydrostatic and kinetic equilibrium, where the dominant cooling process, electron capture on free nucleons, has a sharp decrease with temperature. This dictates that the dominant heating process would be the neutrino absorption of free nucleons followed by the neutrino-electron scattering also contributing to some extent. Thus, an analytical estimate of the mass-loss rate similar to Refs.~\cite{Qian:1996xt,Metzger:2010pp} (also see Ref.~\cite{Desai:2026llu}) for a slowly rotating and unmagnetized neutrino-driven outflow can be given by
\be
\label{eq:masslossneutrino}
\dot{M}_\nu \approx 5 \times 10^{-5} M_\odot\ {\rm s}^{-1} \bigg( \frac{Q_\nu(t)}{10^{54}\ {\rm erg} {\rm MeV}^{-2} {\rm s}^{-1}} \bigg)^{5/3} \bigg( \frac{R_*}{10\ {\rm km}} \bigg)^{5/3} \bigg( \frac{M_*}{1.4 M_\odot} \bigg)^{-2} \bigg( 1 + \varepsilon_{\rm ES} \bigg)^{5/3}\,,
\ee
where $L_\nu(t)$ is the time-dependent neutrino luminosity, $\varepsilon_\nu(t)$ is the average neutrino energy, and the term $(1+\varepsilon_{\rm ES})$ accounts for the correction due to elastic scatterings. Note that the prefactor matches with the \texttt{NRNM} model described in Ref.~\cite{Desai:2026llu}.

\begin{figure}
\includegraphics[width=0.9\textwidth]{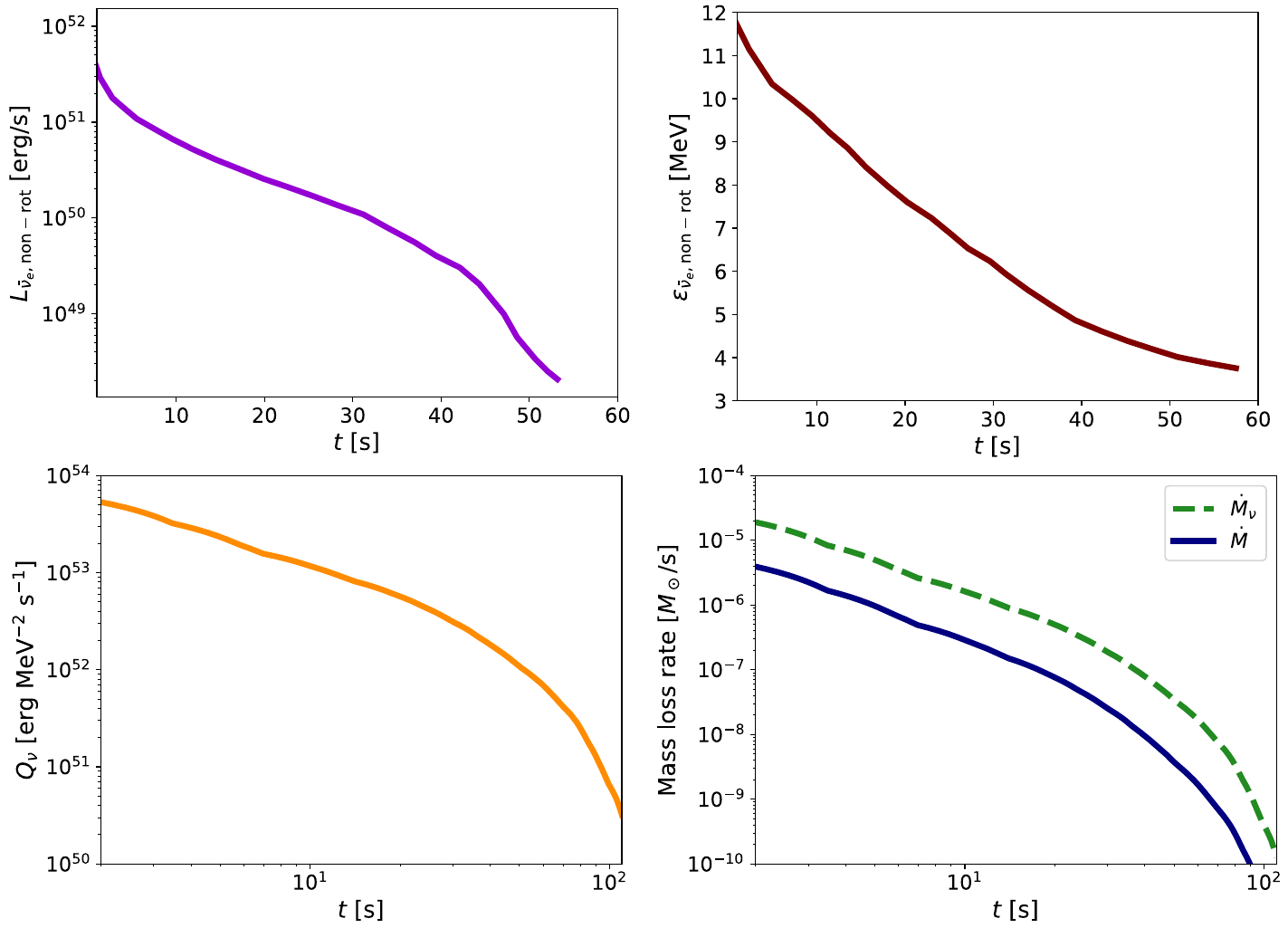}
\caption{\label{appfig:massloss}\emph{Top: }The time evolution of $\bar{\nu}_e$ luminosity ($L_{\bar{\nu}_e}$, \emph{left}) and average energy ($\varepsilon_{\bar{\nu}_e}$, \emph{right}) for a $1.4M_\odot$ PNS from Refs.~\cite{Nakazato:2017ynw,Suwa:2019svl}. \emph{Bottom: }The time evolution of the neutrino heating rate ($Q_\nu$, \emph{left}) estimated from $L_{\bar{\nu}_e}$ and $\varepsilon_{\bar{\nu}_e}$ and the mass loss rate (\emph{right}) for neutrinos $\dot{M}_\nu$ (dashed dark green line) and the total mass loss rate $\dot{M}$ (solid dark blue line). See the text for details.
}
\end{figure}
We use the long-term simulation results from Refs.~\cite{Nakazato:2017ynw,Suwa:2019svl} for the model \texttt{147S}, in which the shock stalling happens at $t=0.3$ s post bounce and the baryonic mass of the PNS $m_b = 1.47 M_\odot$ and the corresponding gravitational mass $M_* = 1.35 M_\odot \sim 1.4 M_*$ consistent with what we consider in the current work. The corresponding time evolution of the electron antineutrino $\bar{\nu}_e$ luminosity ($L_{\bar{\nu}_e}$) and the average energy $\varepsilon_{\bar{\nu}_e} = \langle E_{\bar{\nu}_e} \rangle$ is shown in the \emph{top} panels of Fig.~\ref{appfig:massloss}. Since we are interested in the late-time evolution well after the neutronization burst and the accretion phase, we approximate that $L_{\nu_e} \approx L_{\bar{\nu}_e}$ and $\varepsilon_{\nu_e} \approx \varepsilon_{\bar{\nu}_e}$~\cite{Fiorillo:2023frv,Lucente:2024ngp}. What we are interested in evaluating is the neutrino heating rate $Q_\nu \propto (L_\nu \langle E_\nu^2\rangle)$. This can be estimated easily by using the pinching parameter $\alpha_{\nu_i}$ such that $\langle E_{\bar{\nu}_i}^2 \rangle \approx (2+\alpha_{\nu_i})/(1+\alpha_{\nu_i}) \varepsilon_{\nu_i}^2$. We choose $\alpha_{\nu_e} \approx \alpha_{\bar{\nu}_e} = 2.5$. With these approximations we have $Q_\nu \approx 2 L_{\bar{\nu}_e} \langle E_{\bar{\nu}_e}^2 \rangle$ which is shown in Fig.~\ref{appfig:massloss} \emph{(bottom left)}. We note that for the relevant timescales when CRs can be accelerated by the jet $t\lesssim 20$ s, the neutrino heating rate decreases by a factor of a few. It is important to note that the long-term simulations we use do not include some of the micro-physical aspects like convection and muon cooling. In Ref.~\cite{Fiorillo:2023frv} it was shown that including such effects would reduce the neutrino transparency of the PNS earlier ($t\sim 10$ s) than what is shown in Fig.~\ref{appfig:massloss}. This would then reduce $Q_\nu$ leading to a much steeper decrease in the mass loss rates and as a result an increase in $\sigma_0$. Therefore, the jet would become ultra-relativistic and magnetically dominated even earlier than what we estimate in this work. Thus, our estimate is conservative in this sense. Finally, the simulations do not account for inelastic electron scattering which leads to additional heating. Similar to Refs.~\cite{Qian:1996xt,Metzger:2010pp}, we introduce an order $1$ correction by estimating this as, $\varepsilon_{\rm ES} \approx 0.2 (M_{\rm NS}/1.4\ M_\odot) (R_*/10\ {\rm km})^{-1} (\varepsilon_{\nu,i}/10\ {\rm MeV})^{-1}$.

There are two main conditions we assume for the PNS as far as the analytical estimate of $\dot{M}$ is concerned -- non-rotating and unmagnetized. Both these assumptions modify $\dot{M}$ and in the following we estimate to some extent the corrections introduced to $\dot{M}$ as a result of this. Let us first begin by addressing the aspect of rotating PNS. A rapidly rotating PNS has a slower cooling as compared to the weakly-rotating or non-rotating ones. This is because the rotation typically reduces the interior temperature of the neutron star. Indeed, the neutrino luminosities and mean energy receive an order unity correction a result of rotating PNS. In the absence of dedicated long-term simulations we adopt the stretch factor $\eta_{\rm rot}$ similar to Refs.~\cite{Thompson:2004wi,Metzger:2010pp}, $\eta_{\rm rot} = 2.5$. Therefore, as compared to the non-rotating case, we have the following scaling for the neutrino luminosity and mean energy, $L_{\nu,\rm rot} (t) = \eta_{\rm rot}^{-1} L_{\nu,\rm non-rot}(t/\eta_{\rm rot})$ and $\varepsilon_{\rm rot}(t) = \eta_{\rm rot}^{-1/4} \varepsilon_{\rm non-rot}(t/\eta_{\rm rot})$.

Let us now discuss the effect of magnetized PNS. First of all our initial assumption was that the outflow is spherical and stems from the entire magnetar surface. For a magnetized PNS, the strong magnetic field dictates that the plasma flows along open field lines. Therefore, closed field lines do not contribute to the outflow. Assume a fraction $f_{\rm op}$ of the total surface area of the PNS are threaded by open field lines ($f_{\rm op} \approx A_{\rm op}/(4\pi R_*^2)$) such that the outflow gets scaled by a factor of $f_{\rm op}$, that is, $\dot{M} = f_{\rm op} \dot{M}_\nu$. We can estimate $f_{\rm op}$ by the following. We assume that the last open field line is at the light cylinder radius $R_{\rm LC} = c/\Omega$. This is a reasonable assumption for a highly magnetized $\sigma_0 \gg 1$ scenario~\cite{.} and force-free electrodynamics. The area of the polar cap consisting of open field lines can be roughly estimated as $A_{\rm op} \sim 2 \times 2 \pi R_*^2 \big(1 - \cos \theta_{\rm op} \big)$, where the factor of $2$ stems from the contribution from the two hemispheres and that $\theta_{\rm op}$ is the half-opening angle of the polar cap. Thus, $f_{\rm op} \approx 2 \sin^2 (\theta_{\rm op}/2)$. Now, for a rotating dipole, the last open field line satisfies $\sin^2 \theta_{\rm op} \sim R_*/R_{\rm LC}$, using which we can evaluate $f_{\rm op}$.

Next, we look at a combined effect of rotation and the magnetic fields on the PNS and hence the mass loss rate~\cite{Thompson:2004wi,Metzger:2010pp}. The plasma co-rotates with the magnetic field lines. Along the open field lines, the centrifugal force component provides an outward force, leading to a reduction in the gravitational potential, and therefore, an enhancement in the mass loss rate. We quantify this by $f_{\rm cent}$. The mass loss rate is exponentially sensitive to the effective potential ($\dot{M} \propto \exp \big( - \Phi_{\rm eff}/c_s^2 \big)$, where $\Phi_{\rm eff}$ and $c_s$ are the effective potential and the sound speed respectively) and thus $f_{\rm cent}$ introduces exponential corrections to the mass loss rate. Since the centrifugal term adds a positive contribution to the potential we have $\Phi_{\rm eff} \propto -GM/r + (1/2) \Omega^2 r^2 \sin^2 \theta$, therefore, $f_{\rm cent}\propto \exp(\Omega^2 R_*^2/c_s^2)$.

However, the above phenomenon is self-regularized. This is because as the mass loading is enhanced, $\sigma_0 \propto 1/\dot{M}$ (see Eq.~\ref{eq:sigma0}) decreases leading to the wind becoming less magnetized. A less magnetized wind cannot enforce co-rotation of the surrounding plasma. So for the outflow wind to be highly magnetized, there exists a maximum centrifugal enhancement possible, which we define as $f_{\rm cent}^{\rm max}$. Following Ref.~\cite{Metzger:2007kj} we have $f_{\rm cent}^{\rm max} \approx \exp \big( (\Omega/\Omega_{\rm crit})^\beta \big)$, where $\Omega_{\rm crit}$ is the critical angular velocity where the maximum centrifugal enhancement can occur and $\beta \simeq 1.5$. Now the critical spin period $P_{\rm crit} \propto R_{\perp}/c_s$, where $R_{\perp} \propto R_* \sin \iota$, and the sound speed $c_s \propto (M_*/R_*)^{1/2}$. Thus we have, $P_{\rm crit} \approx 2.1\ {\rm ms}\ (\sin\ \iota) (R_*/10\ {\rm km})^{3/2} (M_*/1.4 M_\odot)^{-1/2}$, where the normalization in front is obtained from a numerical fit~\cite{Metzger:2007kj}. The sonic radius $R_s = \big( GM_*/\Omega^2 \big)^{1/3} = 2.7 \times 10^6\ {\rm cm}\ (M_*/1.4 M_\odot)^{1/3} (P_i/2\ {\rm ms})^{2/3} (1+t/t_{\rm sd})^{1/3}$ determines the enhancement of $\dot{M}$, in this way that, the outflow needs to co-rotate with the PNS up to a radial distance $r \gtrsim R_s$. For our current work, with $\sigma_0 \gg 1$, that distance $r \sim R_{\rm LC} =c/\Omega = 9.5 \times 10^6\ {\rm cm}\ (P_i/2\ {\rm ms}) (1+t/t_{\rm sd})^{1/2}$ and hence $r = R_L > R_s$ which implies that indeed we are in the centrifugal regime such that $f_{\rm cent} \approx f_{\rm cent}^{\rm max}$.

Finally, at late times, when the PNS becomes transparent to neutrinos, the neutrino wind sourcing the mass-loss rate ceases rapidly. The decreasing $\dot{M}$ reaches the Goldreich-Julian (GJ)~\cite{Goldreich:1969sb} mass loss rate for pairs due to the vacuum electric field. Therefore, in this regime we assume the mass-loss rate is governed by the Goldreich-Julian pair extraction rate given by $\dot{M}_{\rm GJ} = Y (m_e/e) \big( 4\pi R_L^2 c \eta_{\rm GJ}|_{R_{\rm LC}} \big)$, where the pair multiplicity is given by $Y$, the GJ charge density at the light cylinder radius is given by $\eta|_{R_{\rm LC}} = \Omega B/(2 \pi c)$. We assume $Y \approx 10^6$. Note that the magnetization at very late times $\sigma_0(t\gg t_{\nu,\rm thin}) \propto 1/Y$, where $t_{\nu,\rm thin}$ is the time at which the PNS becomes transparent to neutrinos. However, the late time has no impact on our results which are mostly governed by the neutrino-driven winds.

We are now in a position to evaluate the mass-loss rate as a result of neutrino-driven winds ($\dot{M}_\nu$) taking into account all the above corrections using Eq.~\eqref{eq:masslossneutrino} and the net mass-loss rate $\dot{M}$ using Eq.~\eqref{eq:masslossrate}. The time evolution of $\dot{M}_\nu$ (dashed dark green) and $\dot{M}$ (solid dark blue) are shown in the \emph{bottom right} panel of Fig.~\ref{appfig:massloss}. We notice that the various corrections discussed above reduces $\dot{M}_\nu$ by a factor of a few.

\section{Regions of particle acceleration}
\label{appsec:reg_partacc}
\begin{figure}
\includegraphics[width=0.9\textwidth]{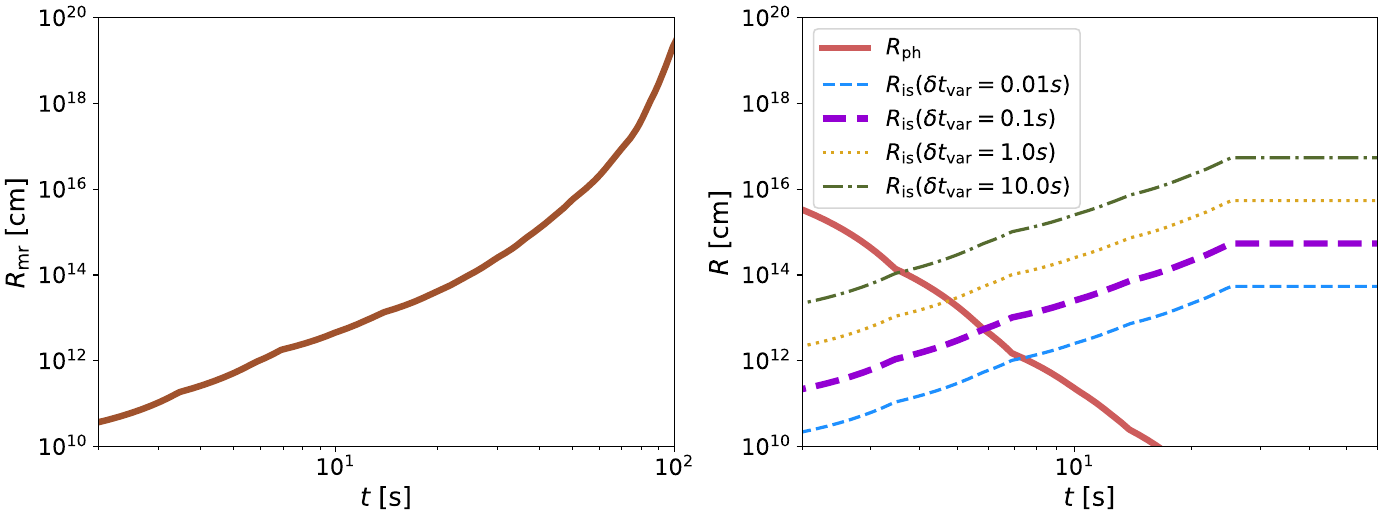}
\caption{\label{appfig:magreconn_intshk}\emph{Top: }The time evolution of the saturation radius for the magnetic reconnection saturation radius $R_{\rm mr}$ \emph{(left)} and the internal shock radius $R_{\rm is}$ \emph{(right)} for different values of $t_j$ along with the photosphere radius $R_{\rm ph}$.
}
\end{figure}
In this section, we present some additional details on the two scenarios of particle acceleration considered in this work - internal shocks and magnetic reconnection.

In Fig.~\ref{appfig:magreconn_intshk} \emph{(left)} we show the time evolution of the saturation radius in the context of magnetic reconnection. Note that $R_{\rm mr} \propto \sigma_0^2(t)P(t)$ and hence roughly follows the evolution of $\sigma_0$ as shown in Fig.~\ref{fig:jet_mod}. In particular, $R_{\rm mr} \sim 10^{14}$ cm at $t \sim 20$ s which is when the peak in $\varepsilon_{\rm CR}^\prime$ occurs.

For the internal shock scenario, we show the time evolution of the photosphere radius $R_{\rm ph}$ and the internal shock radius $R_{\rm is}$ for various values of $t_j$ (the time at which the faster moving shell was launched with respect to the bulk) in Fig.~\ref{appfig:magreconn_intshk} \emph{(right)}. The photosphere radius $\propto L_{k,\rm iso}(t)/\Gamma_j^3(t)$ and decreases with time as moving inwards. On the other hand, $R_{\rm is}\propto \Gamma_j^2(t) t_j$. Since $\Gamma_j$ saturates to $\bar{\Gamma}_j^{\rm is}$ at $t\gtrsim 20$ s, $R_{\rm ph}$ also plateaus. Moreover, with increase in $t_j$ the internal shock radius increases. Most importantly, we demand that for the internal shock scenario to efficiently accelerate CRs, $R_{\rm ph} < R_{\rm is}$, such that the shock is not radiation mediated. In this work, we choose $t_j \sim 1.0$ s for which $R_{\rm is}>R_{\rm ph}$ for $t \gtrsim 5$ s. The peak in $\varepsilon_{\rm CR}^\prime$ indeed occurs for $t>10$ s ensuring that we are in the required regime where the shock is efficient in accelerating the CRs. We also see that given the current choice of our parameters, $t_j \lesssim 1$ s to satisfy the condition of the shock not being radiation mediated.

We can also test the condition for internal shocks to occur by assuming the flow has transitioned to an effectively matter-dominated regime (from a Poynting flux dominated magnetized flow) before internal shocks form. The saturation radius $R_{\rm sat}^{\rm is}$ can then be defined as the radial distance where $\Gamma_j^{\rm is} (r = R_{\rm sat}^{\rm is}) = \bar{\Gamma}_j^{\rm is}$. Moreover, $R_{\rm sat}^{\rm is} = R_0 \bar{\Gamma}_j^{\rm is}$ assuming this fireball-like regime, where $R_0$ is the launch radius of the jet. Therefore, for $\bar{\Gamma}_j^{\rm is} = 300$ as assumed in the main text and typical $R_0 \sim 10^6 - 10^7$ cm, $R_{\rm sat}^{\rm is} \ll R_{\rm is} \gtrsim 10^{13}$ cm for $t_j = 1$ s for all relevant timescales.

Next, we address and clarify another aspect in our setup related to the region of acceleration. From Fig.~\ref{fig:jet_mod}, we observe that at the relevant acceleration phase, the magnetization of the outflow is $\sigma_0 \gtrsim 100$, thus indicating a rarefied outflow. Assuming $p = 2$, the total number of nuclei accelerated as UHECRs in the outflow can be estimated using
$N_{\rm CR} \approx \eta_{\rm CR} E_j/\big(\varepsilon_{\rm min} \ln (\varepsilon_{\rm max}/\varepsilon_{\rm min})\big)$, where the equation is obtained by integrating Eq.~\eqref{eq:crspectra} and neglecting the exponential cut-off for simplicity. For $\eta_{\rm CR} = 10\%$ and $E_j = 3.6 \times 10^{51}\ {\rm erg}$, we have $N_{\rm CR} \sim 3 \times 10^{43}$ (or, $M_{\rm CR, Fe} \sim N_{\rm CR} A m_p \approx 1.5 \times 10^{-12} M_\odot$), which is a small fraction of the mass in the outflow, even for the rarefied case. Now, assuming a fraction $\chi_{\rm Fe}$ of the baryonic outflow is composed of heavy Fe-like nuclei, their availability rate can be given as $\dot{N}_{\rm Fe} \approx \chi_{\rm Fe} \dot{M}/(A m_p)$, where $A = 56$ and $\dot{M} \sim L_j/(\sigma_0 c^2)$ approximately. Therefore, the fraction of available Fe-nuclei that must be accelerated into the UHECR population is
\be
f_{\rm inj}^{\rm Fe} \equiv \frac{\dot{N}_{\rm CR}}{\dot{N}_{\rm Fe}} \simeq \frac{\eta_{\rm CR} \sigma_0 A m_p c^2}{\chi_{\rm Fe} \varepsilon_{\rm min} \ln (\varepsilon_{\rm max}/\varepsilon_{\rm min})} = 8 \times 10^{-4} \bigg( \frac{\eta_{\rm CR}}{0.1} \bigg) \bigg( \frac{\sigma_0}{10^4} \bigg) \bigg( \frac{\chi_{\rm Fe}}{10^{-2}} \bigg)^{-1}\,.
\ee
We clearly see that even for a modest choice of $\chi_{\rm Fe}\sim 1\%$ the required injection fraction of Fe-like nuclei, $f_{\rm inj}^{\rm Fe} \ll 1$. 

\section{Characterizing the uncertainties}
\label{appsec:uncertainty}
\begin{figure}
\includegraphics[width=0.48\textwidth]{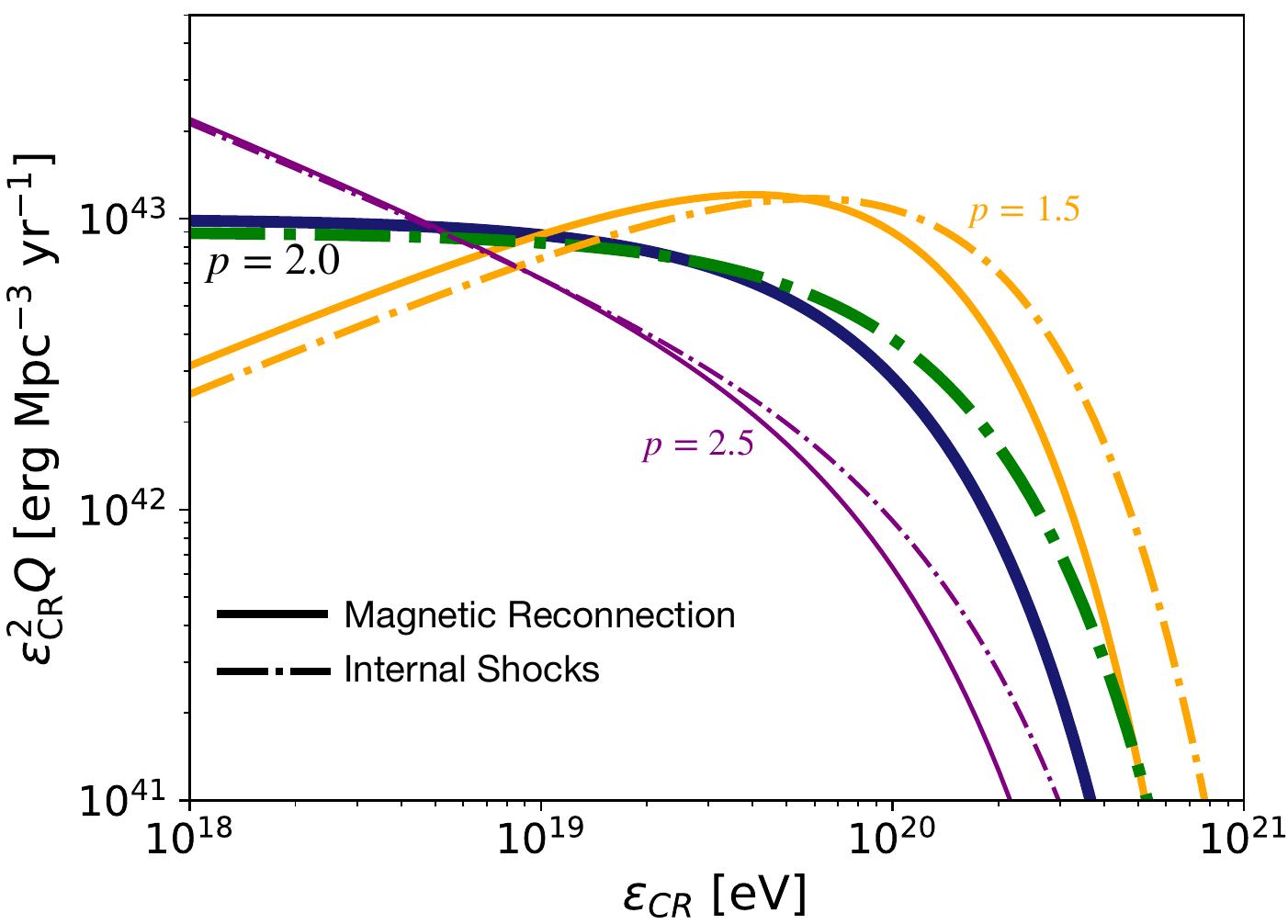}
\caption{\label{appfig:varyp}Same as Fig.~\ref{fig:res_1} but the cases of $p= 1.5$ and $p=2.5$ also shown along with the fiducial $p=2.0$ case.
}
\end{figure}
In this section, we discuss some of the uncertainties relevant for our current work. We mainly focus on two aspects: (a) the parameters for which we get the upper limits for Fig.~\ref{fig:res_1} and (b) the effects when the spectral index of the UHECRs $p$ is varied.

For Fig.~\ref{fig:res_1}, the solid lines denote the reasonable values for the fiducial parameters which we have already discussed in the main text. However, it is worthwhile to also consider optimistic choices for parameters to see how our results change. We achieve this in the following way. To obtain the upper bound, we set $P_i = 1$ ms corresponding to a faster spinning magnetar born from the AIC. A faster spinning magnetar by default has a higher spindown energy and makes our results better. We also set $\eta_{\rm CR} = 0.5$, thus increasing the energy injection to the UHECRs. We keep the magnetic reconnection scenario as is, but for the internal shocks we set $t_j = 10$ s, which still ensure non-radiation mediated shocks. Finally, we choose the upper end of the fiducial rate $\dot{\rho}^{\rm true}_{\rm AIC} = 10^{-6}\ {\rm Mpc}^{-3}{\rm yr}^{-1}$. In this scenario, the maximum energy to which the UHECRs can be accelerated is $\varepsilon^{\prime\rm max}_{\rm CR} = 8.9 \times 10^{19}$ eV and $\varepsilon^{\prime\rm max}_{\rm CR} = 2.8 \times 10^{20}$ eV. This essentially gives us the most optimistic situation possible as far as effective luminosity is concerned. This is shown as the upper limit in Fig.~\ref{fig:res_1} and also corresponds to $L_{\rm eff}^{\rm UHECR} \sim 1.1 \times 10^{37}\ {\rm erg\ s}^{-1}$ in Fig.~\ref{fig:result}. Furthermore, for the effective source number density ($n_{\rm eff}$), for our fiducial case, choosing $\dot{\rho}^{\rm true}_{\rm AIC} = 10^{-6}\ {\rm Mpc}^{-3}{\rm yr}^{-1}$ and $\varepsilon^{\prime\rm max}_{\rm CR} = \rm min[\varepsilon^{\prime\rm max,mr},\varepsilon^{\prime \rm max,is}_{\rm CR}]$ gives us the upper limit of the effective source number density $n_{\rm eff} \sim 1.1\ {\rm Mpc}^{-3}$. This allows us to show the corresponding region of the $L_{\rm eff}^{\rm UHECR} - n_{\rm eff}$ plane that could be populated by UHECRs from AIC of WDs.

Next, we also discuss the uncertainty related to the spectral index associated with the UHECRs as in Eq.~\eqref{eq:crspectra}. The spectral index for the UHECRs as a result of transport or acceleration mechanisms can be uncertain. To encapsulate such variation, we choose $p$ between $1.5$ and $2.5$ and present our results along with the fiducial case ($p=2$ and all other parameters same as the fiducial choices) in Fig.~\ref{appfig:varyp}.
\end{document}